\documentclass[twocolumn,aps,prl,superscriptaddress]{revtex4-2}

\usepackage{amsmath}
\usepackage{graphicx}
\usepackage{MnSymbol}
\usepackage{parskip}
\usepackage{makecell}
\usepackage{booktabs}
\usepackage{multirow}
\usepackage{xcolor}

\makeatletter

\def\section{%
  \@startsection
    {section}%
    {1}%
    {\z@}%
    {0.8cm \@plus 1ex \@minus .2ex}%
    {0.5cm}%
    {\normalfont\small\bfseries\raggedright}%
}

\def\subsection{%
  \@startsection
    {subsection}%
    {2}%
    {\z@}%
    {0.8cm \@plus 1ex \@minus .2ex}%
    {0.5cm}%
    {\normalfont\small\bfseries\raggedright}%
}

\def\subsubsection{%
  \@startsection
    {subsubsection}%
    {3}%
    {\z@}%
    {0.8cm \@plus 1ex \@minus .2ex}%
    {0.5cm}%
    {\normalfont\small\itshape\raggedright}%
}

\makeatother

\begin{document}

\title{Transient fluid removal at soft interfaces: Stationary squeeze-out and dynamic scraping in a block-on-flat contact}

\author{R. Xu}
\affiliation{Peter Gr\"unberg Institute (PGI-1), Forschungszentrum J\"ulich, 52425, J\"ulich, Germany}
\affiliation{State Key Laboratory of Solid Lubrication, Lanzhou Institute of Chemical Physics, Chinese Academy of Sciences, 730000 Lanzhou, China}
\affiliation{MultiscaleConsulting, Wolfshovener str. 2, 52428 J\"ulich, Germany}

\author{T. Tada}
\affiliation{Sumitomo Rubber Industries, Ltd., Material Research \& Development HQ. 2-1-1, Kobe 651-0071, Japan}

\author{D. Ferré Sentis}
\affiliation{Decathlon Footwear industrial community, Lille, France}

\author{B.N.J. Persson}
\affiliation{Peter Gr\"unberg Institute (PGI-1), Forschungszentrum J\"ulich, 52425, J\"ulich, Germany}
\affiliation{State Key Laboratory of Solid Lubrication, Lanzhou Institute of Chemical Physics, Chinese Academy of Sciences, 730000 Lanzhou, China}
\affiliation{MultiscaleConsulting, Wolfshovener str. 2, 52428 J\"ulich, Germany}

\begin{abstract}
Fluid removal from rubber-substrate interfaces is crucial for maintaining friction during walking
and vehicle braking on contaminated surfaces. We study the transient friction of rectangular
rubber blocks sliding against tile and glass surfaces lubricated with water, glycerol, mud, or
silicone grease. Two block configurations with different lengths in the sliding direction were
tested after different stationary waiting times. For water, stationary squeeze-out is nearly
complete before sliding begins. For glycerol, both stationary squeeze-out and sliding-induced
fluid removal are important. For mud and grease, the steady-sliding state is reached after a
sliding distance of the order of the block length, with little dependence on the preceding waiting
time, showing that sliding-induced scraping dominates fluid removal for highly viscous substances.
Dividing the contact into shorter blocks accelerates fluid removal by reducing the drainage
distance and increasing the number of leading edges. Stationary squeeze-out calculations based
on the measured surface roughness are in reasonably good agreement with the glycerol experiments.
The results provide design guidelines for rubber tread blocks with multiscale drainage channels
and sufficient compliance to promote transient fluid removal.
\end{abstract}

\maketitle

\setcounter{page}{1}
\pagenumbering{arabic}




\section{1 Introduction}

\begin{figure}[tbp]
\includegraphics[width=0.47\textwidth,angle=0]{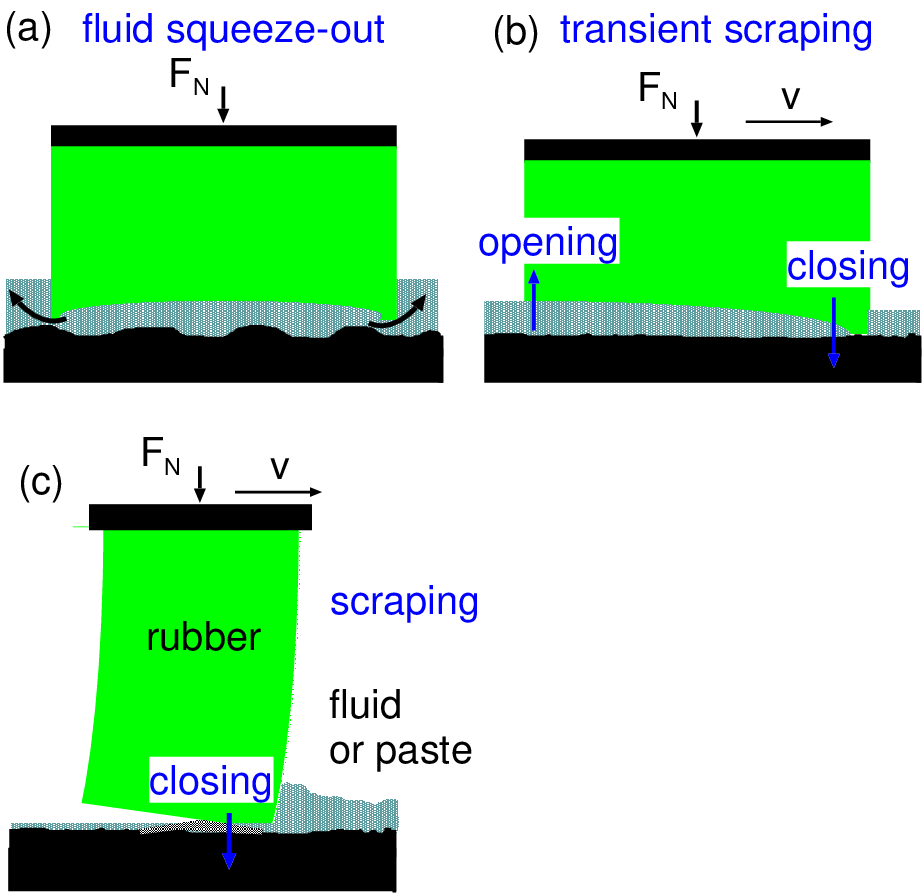}
\caption{The fluid, for example glycerol, mud, or grease, can be removed by
(a) fluid squeeze-out during stationary contact, (b) transient dynamic scraping, or
(c) scraping in which the rubber block bends so that its leading edge contacts the surface
and removes the fluid.
The mechanism in (b) is an elastohydrodynamic effect. During stationary contact, the fluid
pressure initially deforms the rubber surface, as shown in (a). During sliding, a negative
fluid pressure develops near the leading edge and a positive fluid pressure develops near the
trailing edge. This pressure distribution results in fluid removal over a sliding distance of
the order of the length of the nominal contact region in the sliding direction. Depending on
the magnitude of the tangential driving force, the final state may correspond either to a
no-slip state or to steady sliding with elastohydrodynamic deformation of the lower rubber
surface. In (c), a high normal stress may act at the leading edge and remove fluid. In addition,
the inclination of the lower surface of the bent block may generate a negative fluid pressure,
leading to partial or complete closure of the gap.}
\label{Scraping.eps}
\end{figure}

The removal of fluids or mud from rubber-substrate interfaces is crucial in many applications,
including walking and vehicle braking on contaminated surfaces. Most studies of fluid-lubricated
contacts focus on steady-state friction and the Stribeck curve, which describes the friction
coefficient as a function of sliding speed \cite{Japan1,Japan2,Japan3,Japan4,Japan5,Japan6}.
However, many practical contacts involve transient rather than steady sliding.

An important example is the tire-road interaction during braking on fluid-contaminated surfaces.
For vehicles equipped with an anti-lock braking system (ABS), the tire continues to roll during
braking, but its rolling velocity $v_{\rm R}$ is typically $10\%$ lower than the vehicle velocity
$v_{\rm c}$, corresponds to the slip ratio $(v_{\rm c}-v_{\rm R})/v_{\rm c}\approx 0.1$. 
Rubber tread blocks remain within the tire-road
footprint for only a few milliseconds, and the resulting slip distance is typically of the order
of $1 \ {\rm cm}$. The fluid film must therefore be removed rapidly for significant grip
to develop. Similar transient conditions occur during walking or running on contaminated surfaces,
such as mud-covered tracks. In the present study, we investigate fluid removal both during
stationary contact and during subsequent sliding.

Fig.~\ref{Scraping.eps} illustrates three mechanisms by which fluid can be removed from the
interface between a rubber block and a substrate. In Fig.~\ref{Scraping.eps}(a), fluid is removed
by squeeze-out during stationary contact. In Fig.~\ref{Scraping.eps}(b), fluid removal occurs
during sliding through a transient elastohydrodynamic scraping process. The fluid pressure initially
deforms the rubber surface during stationary contact. When sliding begins, a negative fluid
pressure develops near the leading edge and a positive pressure develops near the trailing edge,
promoting fluid removal over a sliding distance of the order of the length of the nominal contact
region in the sliding direction. Depending on the applied tangential force, the final state may
correspond either to no slip or to steady sliding with elastohydrodynamic deformation of the lower
rubber surface.

In Fig.~\ref{Scraping.eps}(c), bending of the rubber block brings its leading edge into contact
with the substrate. A sharp leading edge may generate a high local contact pressure and remove
fluid by direct scraping. In addition, the inclination of the lower surface of the bent block may
produce a negative fluid pressure behind the leading edge, thereby reducing the local interfacial
separation and accelerating fluid removal. Stationary squeeze-out, as illustrated in
Fig.~\ref{Scraping.eps}(a), may also continue during sliding, particularly when sliding begins
shortly after application of the normal load and the fluid film remains relatively thick.

In an earlier study, one of the authors investigated the dependence of lubricated rubber friction on the
stationary contact time and sliding distance for rectangular rubber blocks sliding on smooth
polymer surfaces lubricated with glycerol or grease \cite{JCPsqueeze}. During stationary contact,
the lubricant was removed only slowly from the rubber-polymer interface. During sliding, however,
the fluid was removed much more rapidly. For the grease-lubricated contact, this caused the sliding
motion to stop after a slip distance of only a few times the block length in the sliding direction.
This behavior was attributed to the transient elastohydrodynamic scraping process shown in
Fig.~\ref{Scraping.eps}(b).

In the present work, we study rectangular rubber blocks sliding against tile and glass surfaces
lubricated with water, glycerol, mud, or silicone grease. We examine how fluid removal depends on
the stationary waiting time, sliding distance, fluid properties, and block geometry. We also
compare the experimental results with stationary squeeze-out theory and discuss the implications
for the design of rubber tread blocks that promote rapid fluid removal.


\begin{figure}
\includegraphics[width=0.9\columnwidth]{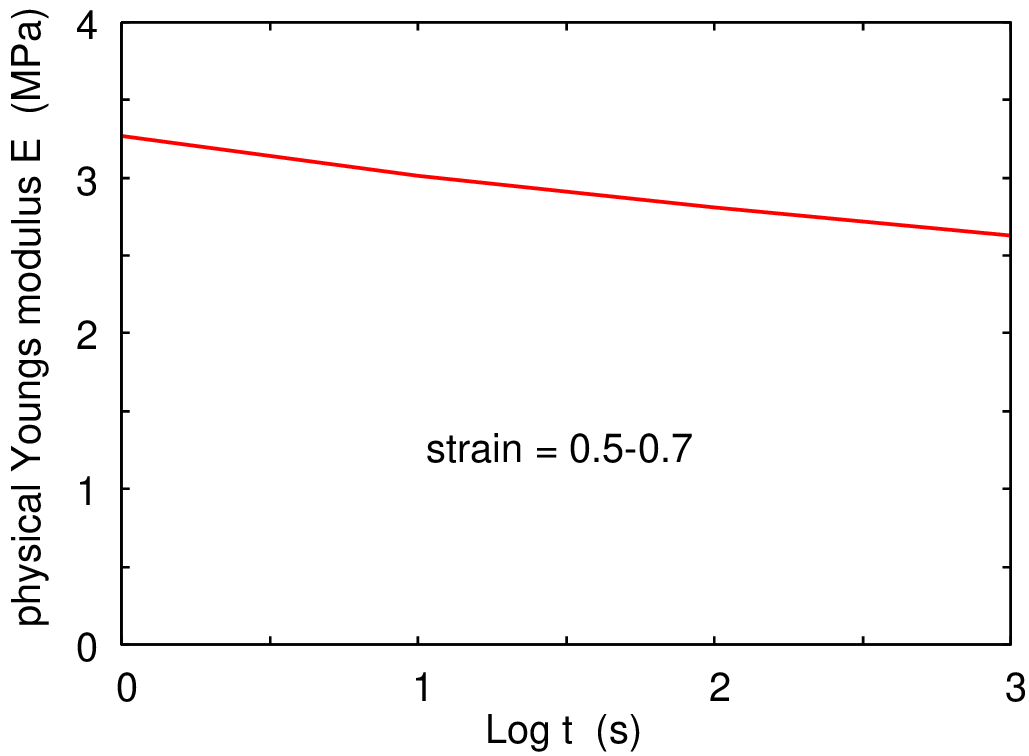}
\caption{\label{1LogTime.2Erelax.JUELICH.eps}
The (physical) relaxation Young's modulus $E(t)$ as a function of time
obtained by loading a rectangular strip of the rubber with a constant force
and measuring the elongation as a function of time.
}
\end{figure}

\section{2 Experimental methods}

Measurements were performed for a shoe-tread rubber compound sliding against tile and glass surfaces. The surface roughness of all contacting surfaces was characterized using a Mitutoyo Surftest SJ-410 portable surface roughness profilometer equipped with a diamond stylus having a tip radius of curvature of $2 \ {\rm \mu m}$ and a stylus-substrate loading force of $0.75 \ {\rm mN}$. The sampling interval (pixel size) was $0.5 \ {\rm \mu m}$, the scan length was $25 \ {\rm mm}$, and the stylus speed was $50 \ {\rm \mu m/s}$. Assuming isotropic surface roughness, the 2D power spectra presented below were calculated from the 1D power spectra obtained by averaging three surface-profile measurements \cite{XuP}.

The shoe-tread rubber compound (supplied by Decathlon France) consisted of polybutadiene rubber (BR) and natural rubber (NR), reinforced with carbon black filler. The viscoelastic modulus of the rubber was characterized using a dynamic mechanical analysis (DMA) instrument and by a simple elongation-relaxation test.

With DMA, the small-strain ($\epsilon = 5 \times 10^{-4}$) shear modulus was measured at a frequency of $1 \ {\rm Hz}$ and a temperature of $T=20^\circ {\rm C}$. The measured value was approximately $G \approx 3.4 \ {\rm MPa}$, corresponding to a small-strain tensile modulus of $E \approx 3G \approx 10 \ {\rm MPa}$.

In the elongation-relaxation test, a rectangular rubber strip was subjected to a constant tensile force, and its elongation was measured as a function of time. The result of this test, performed at $T=20^\circ {\rm C}$, is shown in Fig. \ref{1LogTime.2Erelax.JUELICH.eps}. The figure shows the relaxation Young's modulus $E(t)$ as a function of time. During the measurement, the strain increased from $\epsilon \approx 0.5$ to approximately $0.7$.

Filled rubber compounds exhibit significant strain softening, which probably explains the large
difference between the DMA measurement and the elongation-relaxation experiment. The squeeze-out
problem involves a broad and spatially nonuniform range of strains, extending from relatively small
macroscopic strains to much larger local strains in the asperity contact regions, particularly
during the final stage of squeeze-out when the fluid film becomes very thin. We therefore use a
Young's modulus of $E=5 \ {\rm MPa}$ in the calculations as an approximate intermediate value
between the small-strain DMA modulus and the large-strain modulus obtained from the
elongation-relaxation experiment.

For an unconstrained homogeneous rubber layer, the applied nominal pressure
$p_0=0.16 \ {\rm MPa}$ would correspond to a strain of approximately
$\epsilon\approx p_0/E\approx0.03$. However, this estimate does not represent the actual
compression of the bonded rubber block. The lateral deformation of the rubber is constrained by
its attachment to the rigid support, and the effective compressive stiffness therefore depends on
the block geometry and strain level \cite{Gent1,Gent2,Gent3,Gent4}. The actual macroscopic
compressive strain is consequently expected to be smaller than the simple estimate
$p_0/E$.

\begin{figure}
\includegraphics[width=0.9\columnwidth]{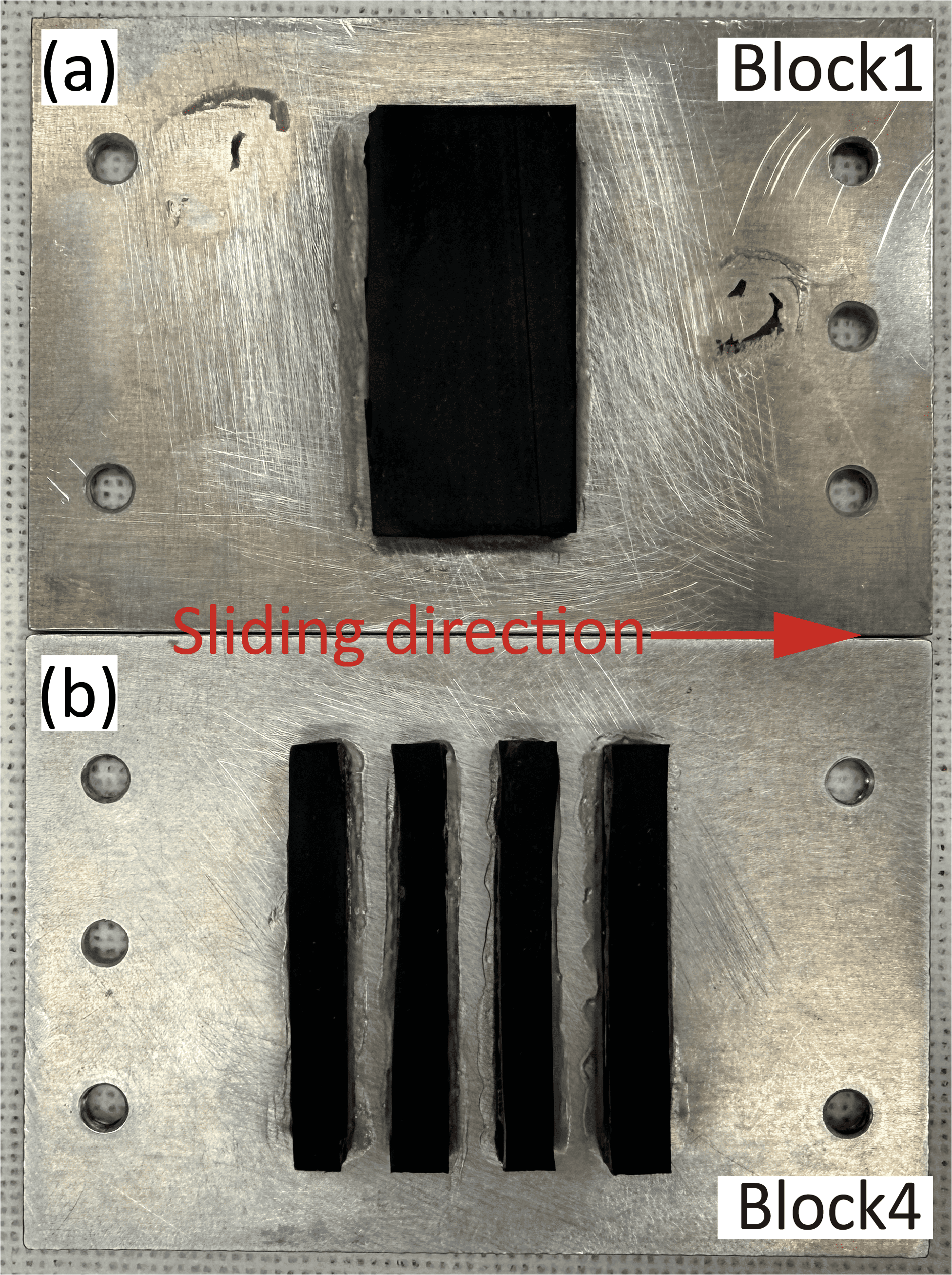}
\caption{\label{PhotoTwoBloksUsed.eps}
Rubber-block configurations used in the present study. All blocks are
$5 \ {\rm cm}$ wide and $0.6 \ {\rm cm}$ high. The block in (a) is
$2.5 \ {\rm cm}$ long in the sliding direction, whereas each of the four blocks in (b) is
$2.5/4=0.625 \ {\rm cm}$ long in the sliding direction.
}
\end{figure}

The rubber specimens were cut from $0.6 \ {\rm cm}$ thick sheets and arranged in the two
configurations shown in Fig.~\ref{PhotoTwoBloksUsed.eps}. All blocks are $5 \ {\rm cm}$ wide
in the direction perpendicular to sliding. The single block shown in Fig.~\ref{PhotoTwoBloksUsed.eps}(a)
is $2.5 \ {\rm cm}$ long in the sliding direction. In the configuration shown in
Fig.~\ref{PhotoTwoBloksUsed.eps}(b), the same total length is divided into four blocks, each
$2.5/4=0.625 \ {\rm cm}$ long in the sliding direction. We refer to the configurations in
Fig.~\ref{PhotoTwoBloksUsed.eps}(a) and (b) as Block1 and Block4, respectively.

The interfaces were lubricated with deionized (DI) water, glycerol with a purity of 99.5\%,
mud, or silicone grease. Fresh glycerol was used for each measurement to avoid a reduction in
viscosity caused by the absorption of water from the atmosphere.
The mud was prepared by gradually adding bentonite powder to water at an approximate
water-to-bentonite powder volume ratio of $4:3$. Bentonite powder strongly absorbs water and
forms a non-Newtonian, shear-thinning suspension. The mixture was allowed to hydrate and was
subsequently stirred thoroughly until a macroscopically homogeneous suspension without visible
dry-powder agglomerates was obtained.
The silicone grease consisted of silicone oil thickened with amorphous fumed silica.

The friction experiments were performed using a linear friction tester for which the normal force could be varied from $F_{\rm N}=31 \ {\rm N}$ to approximately $1000 \ {\rm N}$. The sliding velocity could be varied from $1 \ {\rm \mu m/s}$ to $1 \ {\rm cm/s}$, although most of the waiting-time experiments reported here were performed at $v=1 \ {\rm mm/s}$. During the experiments, the substrate, consisting of either the tile surface or the glass surface, was attached to the machine table. The table was translated by a servo motor through a gearbox. This arrangement enabled precise control of the relative sliding velocity between the specimen and the substrate.

\section{3 Squeeze-out theory}

We review the basic theory of fluid squeeze-out under stationary contact conditions, i.e., in the absence of sliding. Consider first a rectangular elastic block squeezed against a nominally flat substrate in the presence of a fluid. The surfaces exhibit random roughness characterized by the combined surface roughness power spectrum $C(q)$. Because of the surface roughness, non-contact regions will exist at the interface. If the squeezing force is not too large, interconnected non-contact channels may extend from one side of the nominal contact region to the other.

We assume that the fluid is Newtonian and incompressible. Let $u({\bf x},t)$ denote the local surface separation at the position ${\bf x}=(x,y)$ and time $t$, and let ${\bf J}({\bf x},t)$ denote the fluid volume-flow current parallel to the interface. Conservation of fluid volume gives
$${\partial u \over \partial t}+\nabla\cdot{\bf J}=0\eqno(1)$$
where
$${\bf J}=-{u^3\over 12\eta}\nabla p\eqno(2)$$
and $p({\bf x},t)$ is the local fluid pressure.

Equations (1) and (2) describe fluid flow at the interface between contacting solids with rough surfaces. The effects of surface roughness can be eliminated, or integrated out, using a renormalization-group (RG) procedure. In this approach, the surface roughness components are eliminated successively, resulting in a set of RG flow equations that describe how the effective fluid-flow equation evolves as increasingly shorter-wavelength roughness components are removed \cite{PS4}. Equivalent equations were derived by Patir and Cheng using a different approach \cite{Pat1,Pat2}.

If there is a separation of length scales such that the longest roughness wavelength is smaller than the width of the nominal contact region, then, after all roughness components have been integrated out, one obtains
$${\partial \bar u\over\partial t}
-\nabla\cdot\left(
{\bar u^3\phi_{\rm p}(\bar u)\over 12\eta}
\nabla\bar p
\right)=0\eqno(3)$$
where the pressure flow factor $\phi_{\rm p}(\bar u)$ depends on the mean surface separation $\bar u$. In (3), $\bar u$ and $\bar p$ are ensemble-averaged quantities. The pressure flow factor is proportional to the effective fluid-flow conductivity and can be calculated accurately using Bruggeman effective-medium theory, modified such that the contact area percolates at a relative contact area of $A/A_0\approx0.42$, as observed in accurate numerical simulations \cite{Martin}. For discussions of fluid-flow and friction factors in lubricated contacts involving surface roughness, see Refs. \cite{PS1,PS2,PS3,PS4}.

We now consider fluid squeeze-out under an infinitely long rectangular rubber block of width
$w=2a$ in the fluid-flow direction. We introduce a Cartesian coordinate system in the nominal
contact plane, with the origin at the center of the nominal contact region. The $x$-axis is
oriented along the fluid-flow direction, which corresponds to the sliding direction in the
experiments considered below, while the $y$-axis is oriented perpendicular to the sliding
direction. The boundaries of the nominal contact region are therefore located at $x=\pm a$. If the mean interfacial separation $\bar u$ is assumed to depend only on time, (3) reduces to
$${\partial \bar u\over\partial t}
-{\bar u^3\phi_{\rm p}(\bar u)\over 12\eta}
{\partial^2\bar p\over\partial x^2}=0\eqno(4)$$
Assuming that the fluid pressure vanishes at $x=\pm a$, the pressure distribution is
$$\bar p(x,t)=p_{\rm fluid}(t){3\over2}
\left[
1-\left({x\over a}\right)^2
\right]\eqno(5)$$
where $p_{\rm fluid}(t)$ is the spatially averaged fluid pressure. Substituting (5) into (4) gives
$${d\bar u\over dt}
=-{\bar u^3\phi_{\rm p}(\bar u)\over 4\eta a^2}
p_{\rm fluid}(t)\eqno(6)$$

If $p_0$ is the externally applied nominal pressure acting on the upper surface of the rubber block, force balance gives
$$p_{\rm fluid}(t)=p_0-p_{\rm cont}(t)\eqno(7)$$
where $p_{\rm cont}$ is the locally averaged, or ensemble-averaged, asperity contact pressure. The relation between $\bar u$ and $p_{\rm cont}$ can be obtained from Persson contact mechanics theory. At large interfacial separations, this relation takes the form
$$p_{\rm cont}=\beta E^*e^{-\bar u/u_0}\eqno(8)$$
where $E^*=E/(1-\nu^2)$ is the effective elastic modulus, and $\beta$ and $u_0$ are determined by the surface roughness power spectrum. Typically, $u_0\approx h_{\rm rms}$, where $h_{\rm rms}$ is the root-mean-square surface roughness.

The treatment presented above neglects the macroscopic deformation of the elastic block. Since
the fluid pressure is highest along the centerline $x=0$, the lower surface of the block bends
upwards, as illustrated in Fig.~\ref{Scraping.eps}(a). At sufficiently long stationary-contact
times, this bending becomes small and has a negligible influence on the final stages of the
fluid-removal process shown in Fig.~\ref{Scraping.eps}(a).

For perfectly smooth surfaces, $\phi_{\rm p}=1$, $p_{\rm cont}=0$, and $p_{\rm fluid}=p_0$. Equation (6) then becomes
$${d\bar u\over dt}
=-{\bar u^3p_0\over 4\eta a^2}\eqno(9)$$
which can be integrated to give
$${1\over\bar u^2(t)}
-{1\over\bar u^2(0)}
={2tp_0\over\eta w^2}\eqno(10)$$

Fluid squeeze-out is generally a very slow process unless the surface roughness is sufficiently large to provide effective drainage channels. If surface roughness is neglected, the fluid-film thickness is given by (10). In most cases of interest, $1/\bar u(t)\gg1/\bar u(0)$, and the term involving $\bar u(0)$ can therefore be neglected. As an example, consider parameters similar to those used in the experiments described below. For $p_0=0.1\ {\rm MPa}$, $w=1\ {\rm cm}$, and $\eta=1\ {\rm Pa\,s}$, (10) predicts that a time of approximately $t\approx10^9\ {\rm s}$, or about $33$ years, is required to reduce the fluid-film thickness to $\bar u=1\ {\rm nm}$. Accounting for the elastic deformation of the solid walls, as illustrated in Fig. \ref{Scraping.eps}(a), would result in an even longer squeeze-out time.

The treatment above does not include surface and interfacial energies, which may have a crucial influence on fluid squeeze-out once asperity contacts begin to form. For a non-wetting fluid-solid system, the formation of the first asperity contacts may nucleate a dewetting transition, during which fluid is removed not only from the asperity contact regions but also from the surrounding non-contact regions. For a wetting fluid under stationary-contact conditions, a molecularly thin fluid film may remain between the surfaces even in regions conventionally identified as ``contact''. However, if the local asperity contact pressure is sufficiently high, forced dewetting may occur, resulting in the complete removal of the fluid from the asperity contact regions \cite{gly,Paris2}. During sliding, forced wetting may occur, even under conditions for which the fluid would be expelled from a stationary contact \cite{Paris1,Paris2}.


\vskip 0.1cm
\section{4 Fluid removal in rubber-block contacts}

We investigate how the removal of fluid from an interface affects the friction force. This involves
non-stationary sliding, and we denote the ratio between the friction force $F_{\rm f}$ and the
normal force $F_{\rm N}$ by $\mu$, even though this symbol is usually used to denote either the steady-state
kinetic friction coefficient $\mu_{\rm k}$ or the break-loose, or static, friction coefficient $\mu_{\rm s}$.
Thus, $\mu(t)=F_x(t)/F_{\rm N}$ generally depends on time.

Using the Block1 and Block4 configurations described in Sec.~2, we now present the experimental
results for fluid removal at rubber-tile and rubber-glass interfaces.

\subsection{4.1 Experimental results and discussion}

\begin{figure}
\includegraphics[width=1.0\columnwidth]{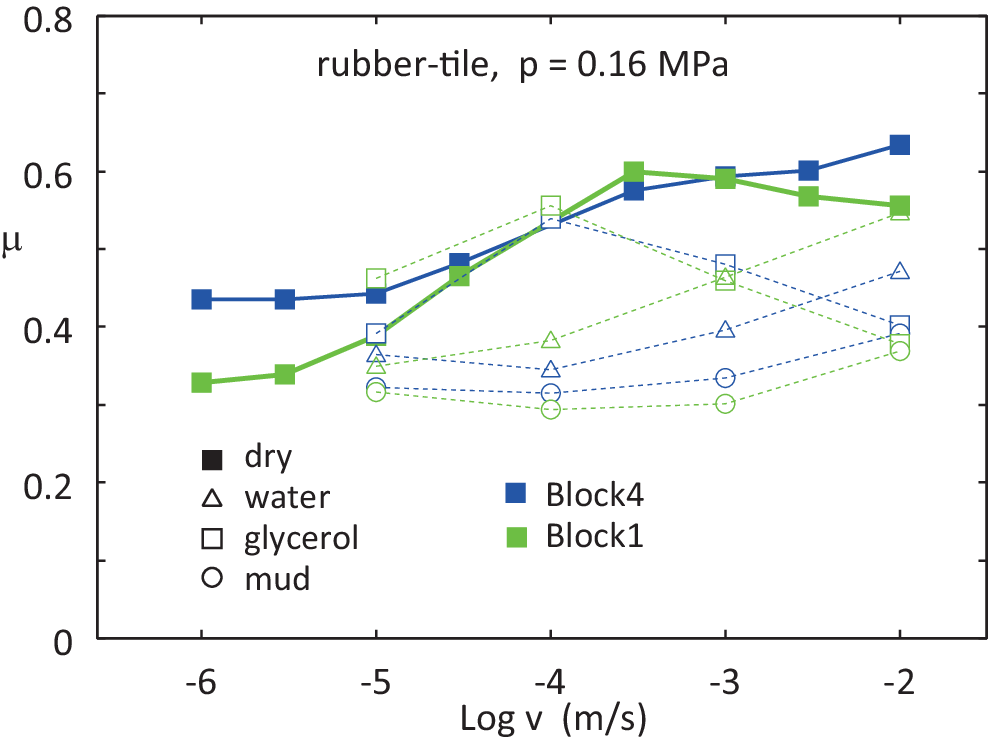}
\caption{\label{BLOCK.1logv.2mu.mastercurve.all.combi.eps}
Friction master curves on the tile surface for Block1 (green symbols) and Block4 (blue symbols).
The filled squares correspond to dry surfaces, whereas the open symbols correspond to lubrication
with water (triangles), glycerol (squares), and mud (circles). The nominal contact pressure is
$p_0=0.16 \ {\rm MPa}$.
}
\end{figure}

Fig.~\ref{BLOCK.1logv.2mu.mastercurve.all.combi.eps} shows the steady-state friction coefficient as a function
of sliding speed for the tile substrate. The filled squares show the dry friction for Block1
(green) and Block4 (blue), whereas the open symbols correspond to contacts lubricated with
DI water (triangles), glycerol (squares), and mud (circles). The dry friction master curves are
similar, although Block4 exhibits slightly higher friction over most of the investigated
sliding-speed range. In glycerol, the friction at sliding speeds up to
$0.1 \ {\rm mm/s}$ is nearly the same as under dry conditions but decreases rapidly at higher
sliding speeds. In water and mud, the friction coefficient is lower than under dry conditions
over the entire investigated sliding-speed range.

The decrease in friction in glycerol at higher sliding speeds is similar to that observed
previously for other systems lubricated with glycerol \cite{gly}. This suggests that the asperity
contact regions may be effectively dry at sliding speeds up to $0.1 \ {\rm mm/s}$ and that
hydrodynamic lift is negligible within this velocity range. At higher sliding speeds,
hydrodynamic effects increase the interfacial separation and reduce the friction.

\begin{figure}
\includegraphics[width=1.0\columnwidth]{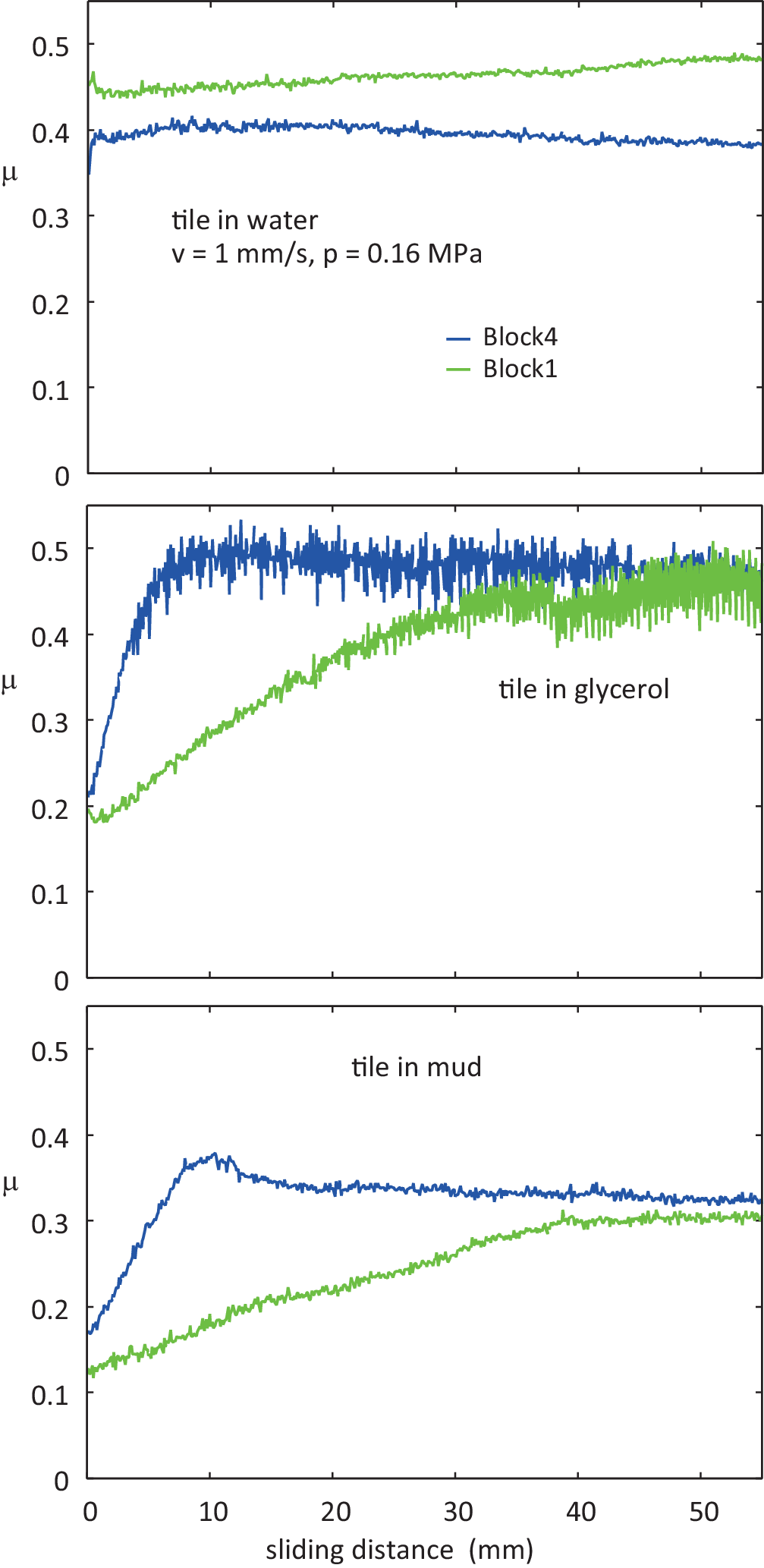}
\caption{\label{BLOCK.1x.2mu.tile.water.glycerol.mud.4Blue.1Green.eps}
The friction coefficient $\mu=F_x/F_z$ as a function of sliding distance for the tile surface
lubricated with (a) water, (b) glycerol, and (c) mud. The blue curves correspond to Block4,
and the green curves correspond to Block1. The sliding speed is $v=1 \ {\rm mm/s}$, and the
nominal pressure is $p_0=0.16 \ {\rm MPa}$.
}
\end{figure}

Fig.~\ref{BLOCK.1x.2mu.tile.water.glycerol.mud.4Blue.1Green.eps} shows the friction coefficient
for Block4 and Block1 as a function of sliding distance for the tile substrate lubricated with
(a) water, (b) glycerol, and (c) mud. In these experiments, the rubber blocks were brought into
contact with the lubricated tile surface and maintained in stationary contact for approximately
$10 \ {\rm s}$. The substrate was then moved at $v=1 \ {\rm mm/s}$.

For water, the sliding friction reaches its steady-state value immediately after the onset of
sliding. Because of the low viscosity of water, squeeze-out is essentially complete during the
$10 \ {\rm s}$ stationary waiting period. This is not the case for the surfaces covered with
glycerol or mud, for which the initial friction is approximately one-half of the steady-state
friction.

For glycerol and mud, the friction coefficient increases approximately linearly with sliding
distance. The corresponding slopes are $0.06$ and $0.009 \ {\rm mm}^{-1}$ in glycerol and
$0.02$ and $0.005 \ {\rm mm}^{-1}$ in mud for Block4 and Block1, respectively. The larger
initial slopes and shorter sliding distances required to reach the steady state show that fluid
removal during the initial stage of sliding is more effective for Block4 than for Block1.

Although the two configurations have the same total nominal contact area, each individual block
in Block4 has only one quarter of the nominal contact area of Block1. Dividing the contact into
four shorter blocks reduces the characteristic distance that the fluid must travel to reach a free
edge and increases the number of free edges available for drainage. During sliding, Block4 also
has four leading edges, whereas Block1 has only one. This increases the number of locations at
which sliding-induced scraping can occur.

\begin{figure}
\includegraphics[width=1.0\columnwidth]{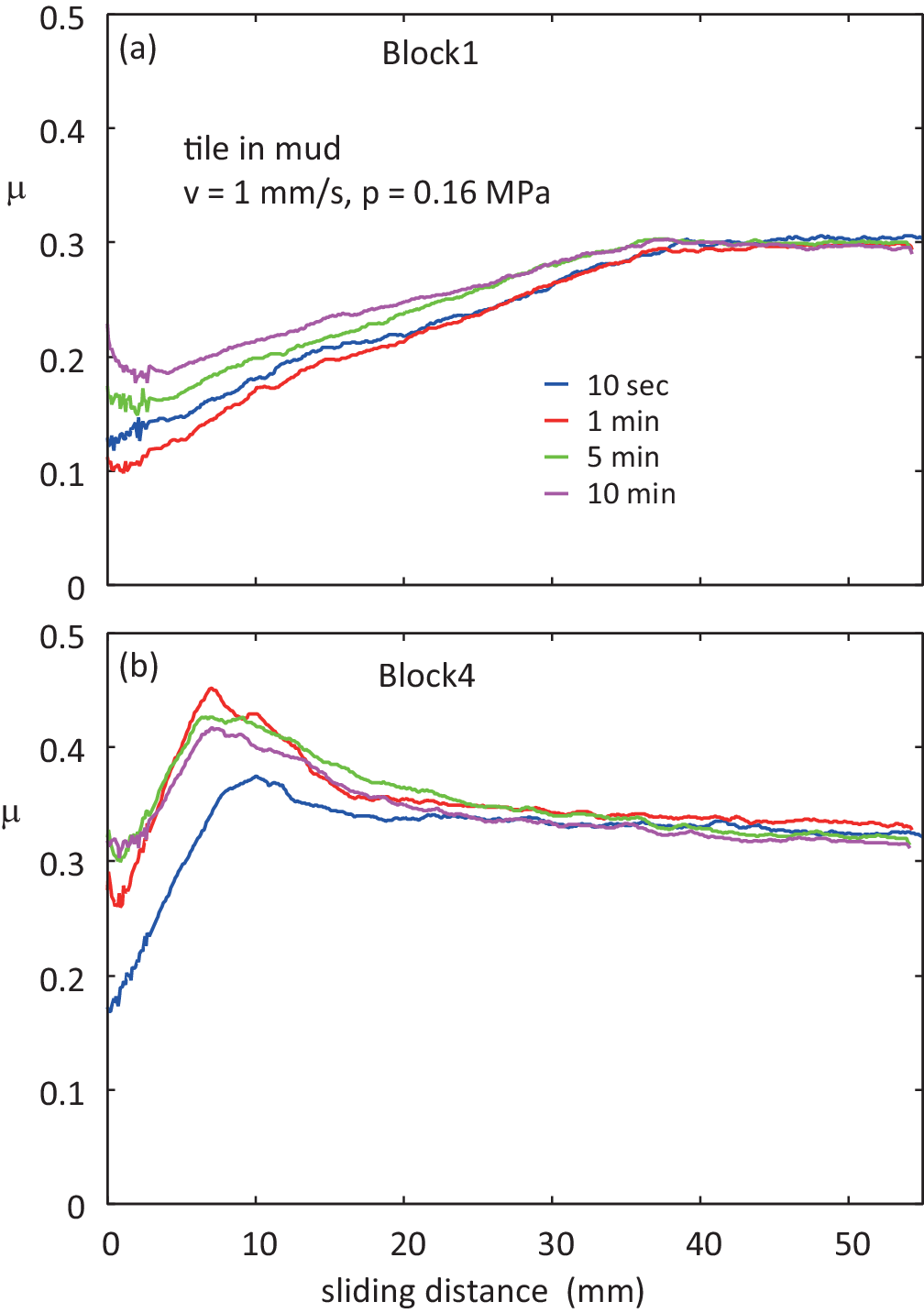}
\caption{\label{BLOCK.1x.2mu.tile.mud.1.and.4.eps}
The friction coefficient as a function of sliding distance for (a) Block1 and
(b) Block4 sliding in mud on the tile surface. The blue, red, green, and purple curves
correspond to waiting times before the onset of sliding of $10 \ {\rm s}$, $1 \ {\rm min}$,
$5 \ {\rm min}$, and $10 \ {\rm min}$, respectively. The sliding speed is
$v=1 \ {\rm mm/s}$, and the nominal pressure is $p_0=0.16 \ {\rm MPa}$.
}
\end{figure}

Fig.~\ref{BLOCK.1x.2mu.tile.mud.1.and.4.eps} shows the friction coefficient as a function of
sliding distance for (a) Block1 and (b) Block4 sliding in mud on the tile surface. The blue, red,
green, and purple curves correspond to waiting periods of $10 \ {\rm s}$, $1 \ {\rm min}$,
$5 \ {\rm min}$, and $10 \ {\rm min}$, respectively. After each waiting period, the substrate
was moved at a sliding speed of $v=1 \ {\rm mm/s}$ over a sliding distance of
$5.5 \ {\rm cm}$.

The friction coefficient at the onset of sliding increases continuously with increasing waiting
time, although the increase between waiting times of $5$ and $10 \ {\rm min}$ is relatively
small, especially for Block4. During sliding, the friction coefficient increases toward nearly
the same steady-state value for all waiting times. This steady-state value is generally higher
than the friction coefficient at the onset of sliding, showing that part of the fluid initially
remaining within the contact is removed during sliding. The final steady state is reached after
a sliding distance of approximately $35 \ {\rm mm}$ for Block1 and $10 \ {\rm mm}$ for
Block4, nearly independently of the preceding waiting time.

At $v=1 \ {\rm mm/s}$, these sliding distances correspond to sliding times of approximately
$35 \ {\rm s}$ and $10 \ {\rm s}$, respectively. These times are much shorter than the
stationary contact times required to produce a comparable increase in friction through the
squeeze-out process shown in Fig.~\ref{Scraping.eps}(a). Thus, {\it rapid fluid removal from the interface between a rubber block and a substrate generally requires a sliding distance of the order of the block dimension in the sliding direction and occurs primarily through the scraping processes shown in Fig.~\ref{Scraping.eps}(b) and (c).}

\begin{figure}
\includegraphics[width=1.0\columnwidth]{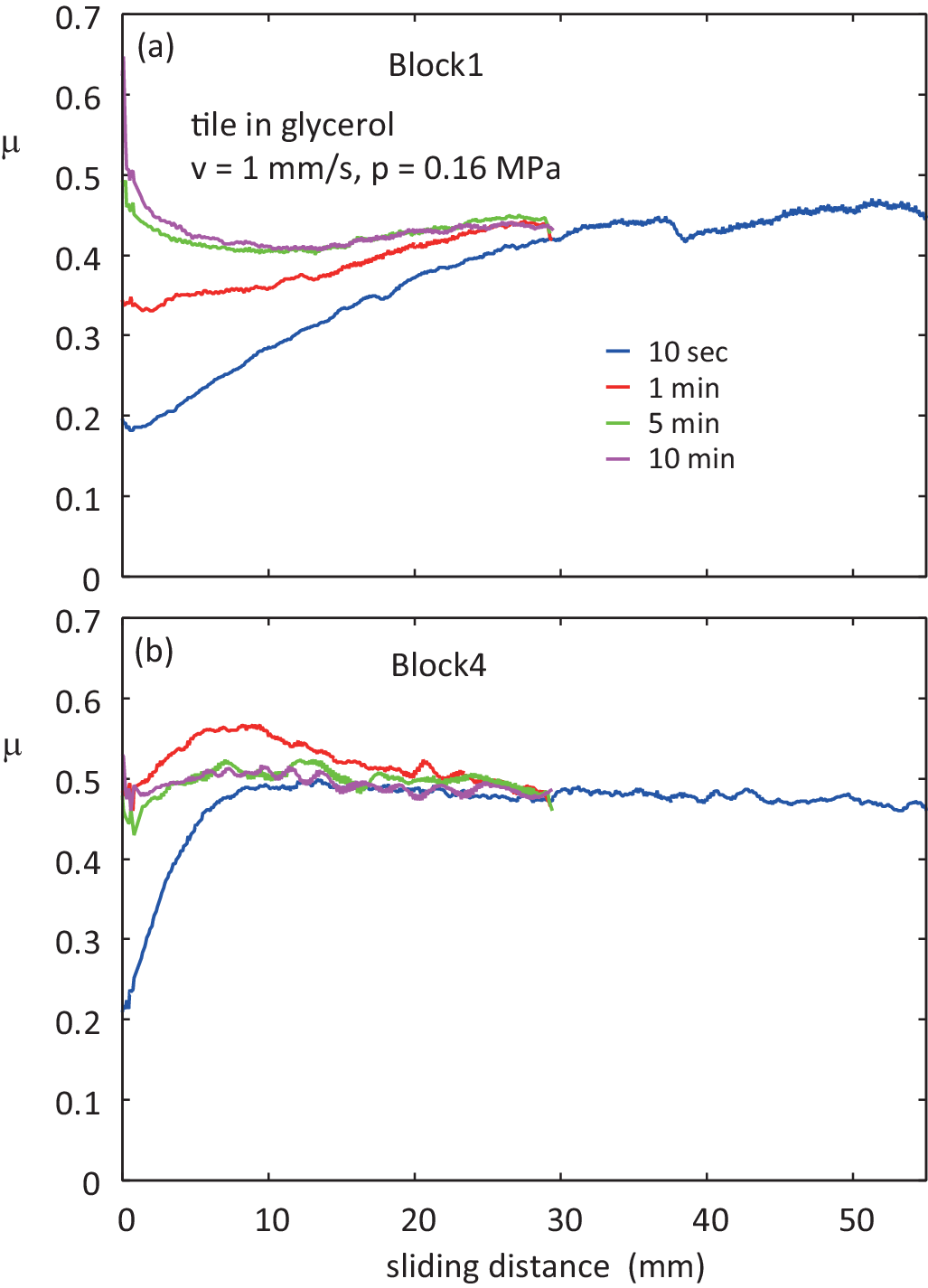}
\caption{\label{BLOCK.1distance.mu.tile.glycerol.Block1And4.eps}
The friction coefficient as a function of sliding distance for (a) Block1 and
(b) Block4 sliding in glycerol on the tile surface. The blue, red, green, and purple curves
correspond to waiting times before the onset of sliding of $10 \ {\rm s}$, $1 \ {\rm min}$,
$5 \ {\rm min}$, and $10 \ {\rm min}$, respectively. The sliding speed is
$v=1 \ {\rm mm/s}$, and the nominal pressure is $p_0=0.16 \ {\rm MPa}$.
}
\end{figure}

Fig.~\ref{BLOCK.1distance.mu.tile.glycerol.Block1And4.eps} shows results similar to those in
Fig.~\ref{BLOCK.1x.2mu.tile.mud.1.and.4.eps}, but for the tile surface lubricated with glycerol.
In this case, fluid squeeze-out during stationary contact is generally more effective than for
mud for both block configurations. After contact times of $5$ and $10 \ {\rm min}$, the friction
at the onset of sliding differs only slightly and is already comparable to, or greater than, the
steady-state friction.

\begin{figure}
\includegraphics[width=0.6\columnwidth]{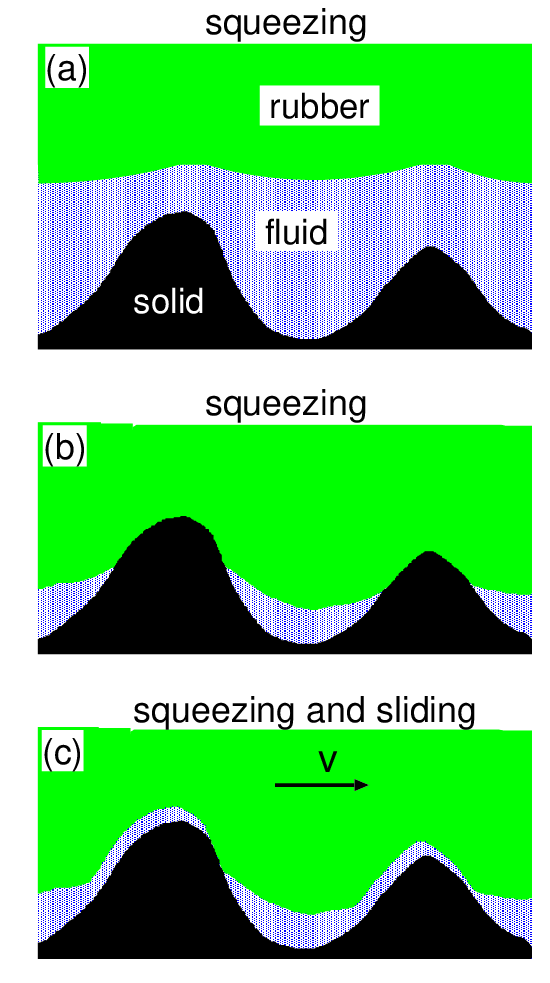}
\caption{\label{AsperityLubrication.eps}
Asperity contact in a fluid. Transition from (a) to (b): if the squeeze-out time is sufficiently
long or the fluid viscosity is sufficiently low, the fluid can be removed from the asperity contact
regions. (c) If, after direct solid-solid contact has been established, the block is slid relative
to the substrate, elastohydrodynamic effects at the asperity scale may generate a fluid film
separating the solids, thereby reducing the friction force. This asperity-lubrication effect exists
even for a rectangular rubber block whose lower surface is parallel to the substrate.
}
\end{figure}

For Block1, the friction at the onset of sliding after contact times of $5$ and
$10 \ {\rm min}$ exceeds the final steady-state value. We attribute this behavior to
hydrodynamic effects during sliding that slightly increase the interfacial separation.
Consequently, the real contact area after a long stationary contact time can be larger than that
in the final steady-sliding state. Even for rectangular blocks that are not macroscopically tilted
relative to the substrate, elastohydrodynamic effects at the asperity scale may reduce the
friction, as illustrated in Fig.~\ref{AsperityLubrication.eps}.

\begin{figure}
\includegraphics[width=1.0\columnwidth]{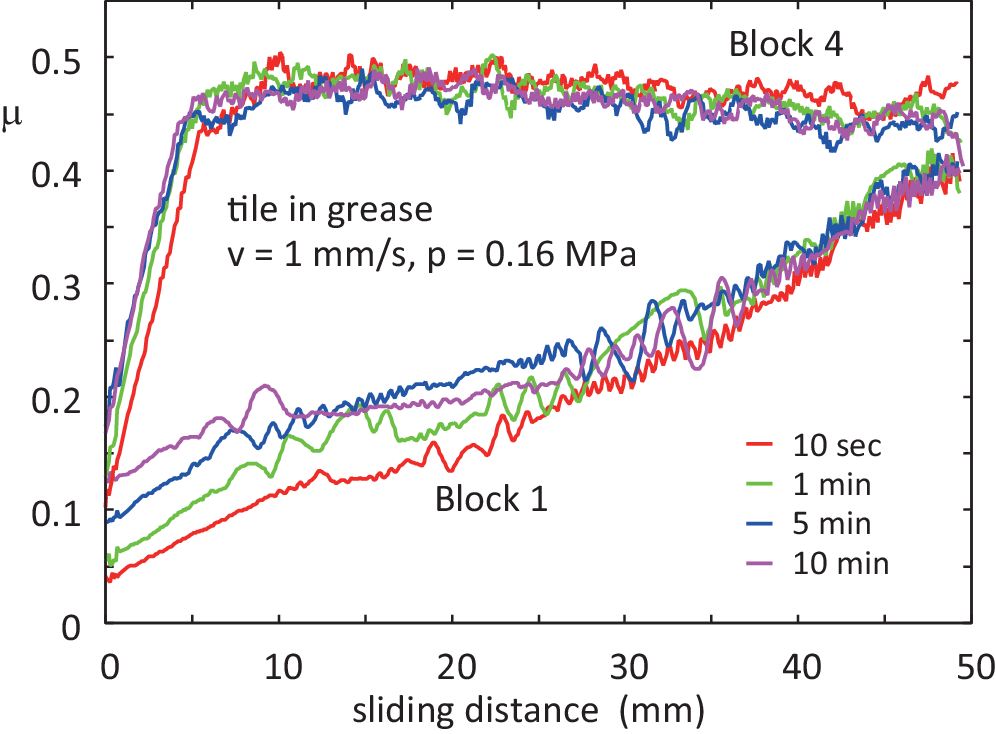}
\caption{\label{BLOCK.1slidingDistance.2mu.tile.grease.eps}
The friction coefficient as a function of sliding distance for Block1 and Block4
sliding in grease against the tile surface. The red, green, blue, and purple curves correspond to
waiting times before the onset of sliding of $10 \ {\rm s}$, $1 \ {\rm min}$,
$5 \ {\rm min}$, and $10 \ {\rm min}$, respectively. The sliding speed is
$v=1 \ {\rm mm/s}$, and the nominal pressure is $p_0=0.16 \ {\rm MPa}$.
}
\end{figure}

Fig.~\ref{BLOCK.1slidingDistance.2mu.tile.grease.eps} shows results similar to those in
Fig.~\ref{BLOCK.1x.2mu.tile.mud.1.and.4.eps}, but for the tile surface lubricated with grease.
Block4 is much more effective than Block1 in removing the lubricant film, as the final
steady-state friction is reached over a much shorter sliding distance. Increasing the stationary
contact time from $10 \ {\rm s}$ to $600 \ {\rm s}$ increases the break-loose friction
coefficient only from approximately $0.035$ to $0.13$. By contrast, during sliding, the friction
coefficient increases from approximately $0.035$ to its steady-state value of approximately
$0.45$ over a sliding distance of $5 \ {\rm cm}$, corresponding to only $50 \ {\rm s}$ at
the imposed sliding speed. Thus, even for Block1, grease removal occurs mainly during sliding.
Although the longest stationary waiting time is much longer than the sliding time, the friction
at the onset of sliding remains far below the final steady-state value.

\begin{figure}
\includegraphics[width=1.0\columnwidth]{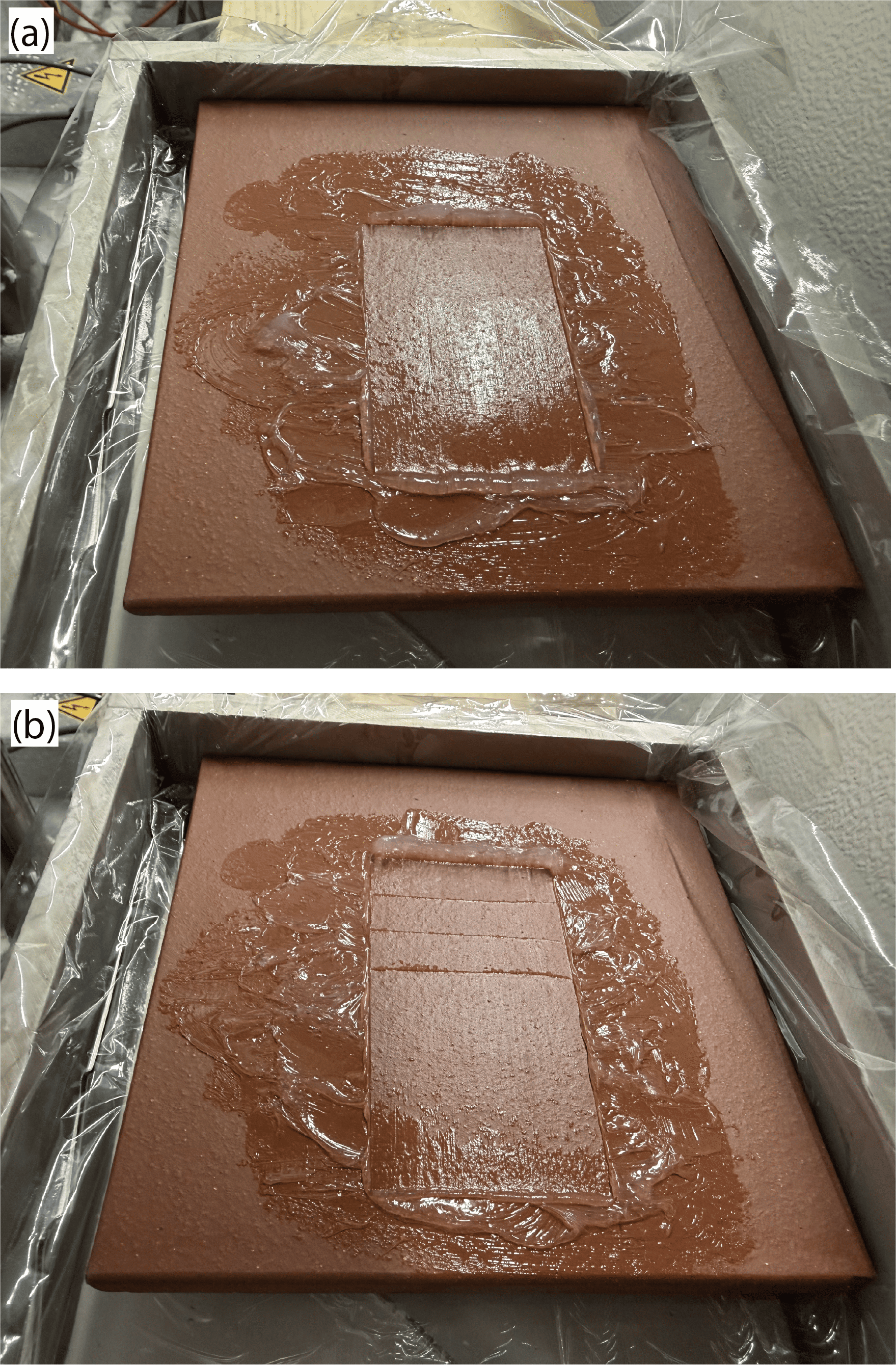}
\caption{\label{BLOCKgrease1Grease2.eps}
Optical images of the tile surface after (a) Block1 and (b) Block4 had been slid over the
grease-covered tile surface.
}
\end{figure}

Fig.~\ref{BLOCKgrease1Grease2.eps} shows the tile surfaces after the sliding experiments presented
in Fig.~\ref{BLOCK.1slidingDistance.2mu.tile.grease.eps}, performed with (a) Block1 and
(b) Block4. Assuming that the initial grease layers had approximately the same thickness, the
shorter shiny region within the sliding track and the smaller amount of residual grease observed
after the Block4 experiment indicate more rapid grease removal. This is consistent with the faster
increase in friction shown in Fig.~\ref{BLOCK.1slidingDistance.2mu.tile.grease.eps}.

The much faster removal of grease by Block4 may result from several geometrical effects. First,
the shorter length of each individual block reduces the distance over which the fluid must be
transported. Second, the four leading edges of Block4 increase the number of locations at which
sliding-induced scraping can occur. The shorter blocks are also less strongly constrained by the
rigid support. For Block1, the characteristic length in the sliding direction is
$2.5 \ {\rm cm}$, which is much larger than the rubber thickness
$d=0.6 \ {\rm cm}$. By contrast, the length of each individual block in Block4 is
$0.625 \ {\rm cm}$, which is comparable to $d$. The individual blocks in Block4 can therefore
bend or tilt more readily, enhancing the scraping process illustrated in
Fig.~\ref{Scraping.eps}(c). Their lower surfaces can also respond more readily to fluid-pressure
variations in the sliding direction, potentially enhancing the transient elastohydrodynamic
scraping process illustrated in Fig.~\ref{Scraping.eps}(b)
\cite{Gent1,Gent2,Gent3,Gent4}.

\begin{figure}
\includegraphics[width=1.0\columnwidth]{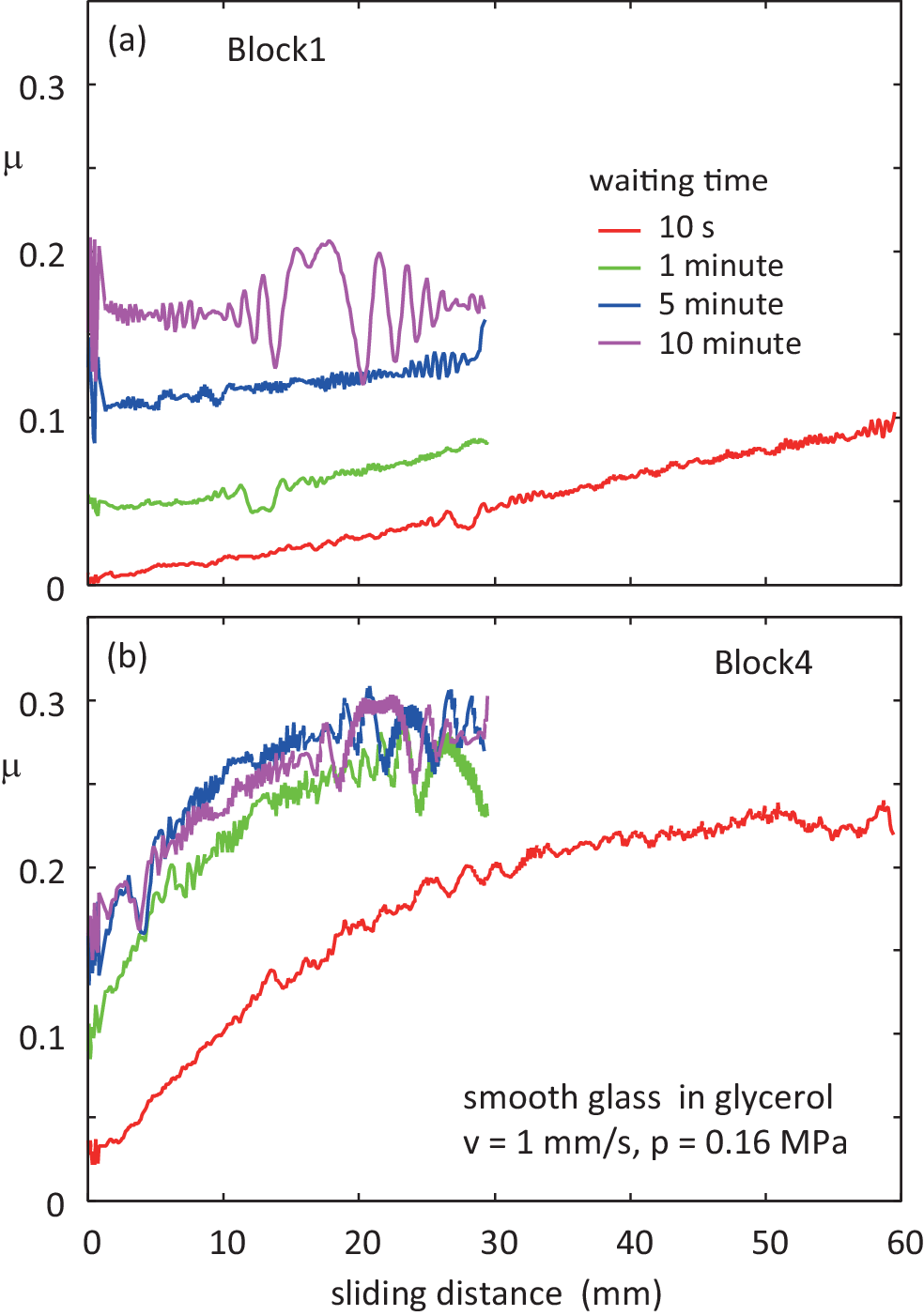}
\caption{\label{BLOCK.no.1.Block.no.4.smoothGLASSglycerol.eps}
The friction coefficient as a function of sliding distance for (a) Block1
and (b) Block4 sliding in glycerol against the smooth glass surface. The red, green, blue,
and purple curves correspond to waiting times before the onset of sliding of $10 \ {\rm s}$,
$1 \ {\rm min}$, $5 \ {\rm min}$, and $10 \ {\rm min}$, respectively.
The sliding speed is $v=1 \ {\rm mm/s}$, and the nominal pressure is
$p_0=0.16 \ {\rm MPa}$.
}
\end{figure}

Fig.~\ref{BLOCK.no.1.Block.no.4.smoothGLASSglycerol.eps} shows the friction coefficient at the
sliding speed $v=1 \ {\rm mm/s}$ as a function of sliding distance for (a) Block1 and
(b) Block4 sliding on a smooth glass surface lubricated with glycerol. The red, green, blue,
and purple curves correspond to waiting times of $10 \ {\rm s}$, $1 \ {\rm min}$,
$5 \ {\rm min}$, and $10 \ {\rm min}$, respectively.

For Block1 and the shortest contact time, the friction force increases approximately linearly
with sliding distance $x$, starting from nearly zero friction at $x=0$. This implies that, after
a squeeze-out time of $10 \ {\rm s}$, the surfaces remain separated nearly everywhere by a fluid
film, resulting in negligible friction at the low sliding speed considered here.

After a contact time of $1 \ {\rm min}$, the friction at the onset of sliding for Block1 is
approximately $25\%$ of the apparent steady-state value. After a contact time of approximately
$10 \ {\rm min}$, fluid squeeze-out is essentially complete before the onset of sliding, and the
friction force becomes nearly independent of sliding distance. The pronounced oscillations in
the friction force after the $10 \ {\rm min}$ contact time are attributed to stick-slip, which is
expected in the present case because the friction decreases with increasing sliding speed for
$v>1 \ {\rm mm/s}$ (not shown).

The smooth-glass results can be interpreted within the same framework as the tile results. In this
case, fluid squeeze-out is governed mainly by the roughness of the rubber surface because the glass
surface is very smooth. Compared with the tile surface lubricated with glycerol
[Fig.~\ref{BLOCK.1distance.mu.tile.glycerol.Block1And4.eps}(a)], for which the friction coefficient
at the onset of sliding is approximately $\mu\approx0.2$ after a waiting time of
$10 \ {\rm s}$, the friction on glass starts from nearly zero. This difference in squeeze-out
efficiency is attributed to the much smaller combined roughness of the rubber-glass interface
than of the rubber-tile interface.

In the Appendix, we present additional results for fluid squeeze-out between Block1 or Block4
and smooth or sandblasted glass surfaces covered by biofilms. Rubber friction on biofilm-covered
surfaces is relevant to robotic hull-cleaning systems \cite{robot}.

\subsection{4.2 Comparison with stationary squeeze-out theory}

\begin{figure}
\includegraphics[width=1.0\columnwidth]{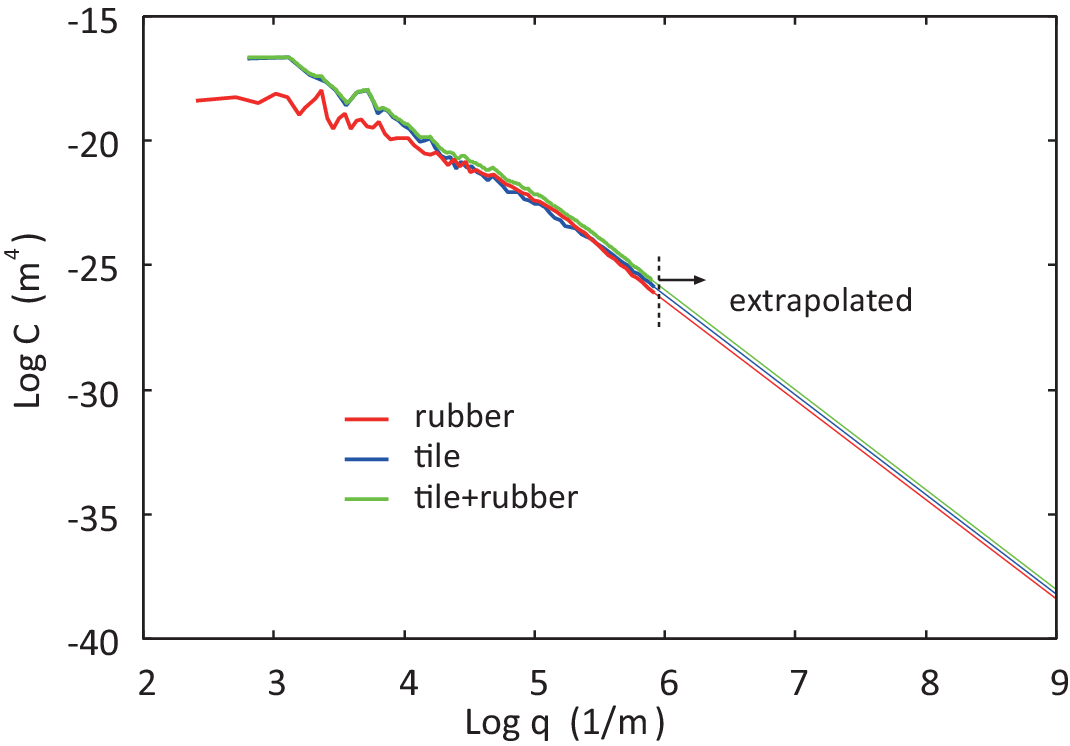}
\caption{\label{BLOCK.1logq.2logC.all.eps}
Surface roughness power spectra as a function of wavenumber on a log-log scale
for the rubber surface (red) and the tile surface (blue). The green curve is the sum of
the power spectra of the two surfaces. The thick curves are obtained from the measured
surface topographies, whereas the thin curves represent linear extrapolations.
}
\end{figure}

\begin{figure}
\includegraphics[width=1.0\columnwidth]{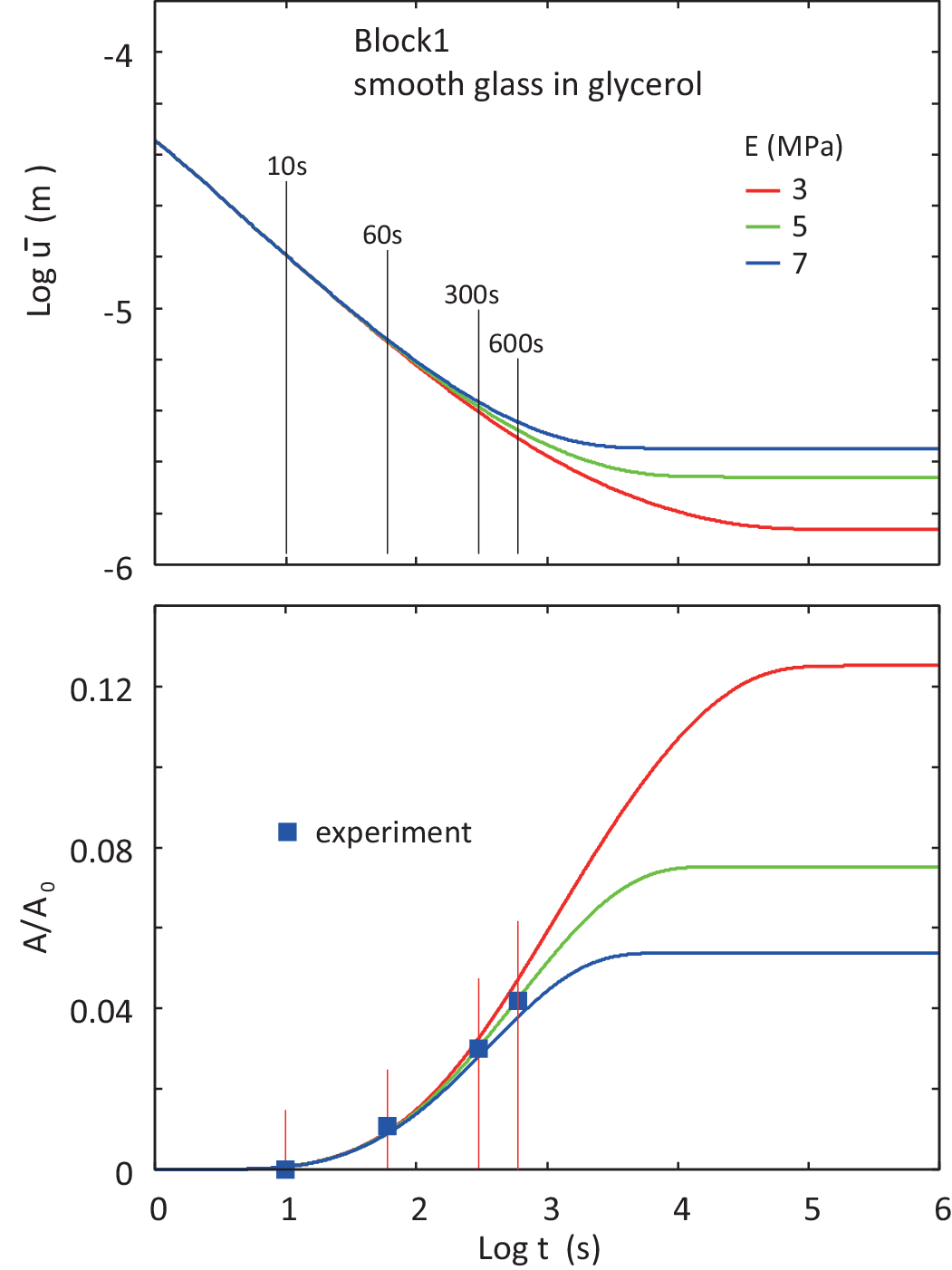}
\caption{\label{BLOCK.1logt.2logu.RubberGlass.3.5.7MPa.eps}
The calculated mean surface separation (a) and relative contact area (b) as functions
of squeeze-out time for Block1 squeezed against a flat surface, here denoted as glass,
in glycerol with viscosity $\eta=1.4 \ {\rm Pa\,s}$. The red, green, and blue curves correspond
to rubber Young's moduli of $E=3 \ {\rm MPa}$, $5 \ {\rm MPa}$, and
$7 \ {\rm MPa}$, respectively. The nominal pressure is $p_0=0.16 \ {\rm MPa}$,
and the initial surface separation is $100 \ {\rm \mu m}$. The squares in (b) show the
relative contact areas obtained from (11), using the experimental friction coefficients
at the onset of sliding for different waiting times from
Fig. \ref{BLOCK.no.1.Block.no.4.smoothGLASSglycerol.eps} and a frictional shear stress
of $\sigma_{\rm f}=0.8 \ {\rm MPa}$.
}
\end{figure}

\begin{figure}
\includegraphics[width=1.0\columnwidth]{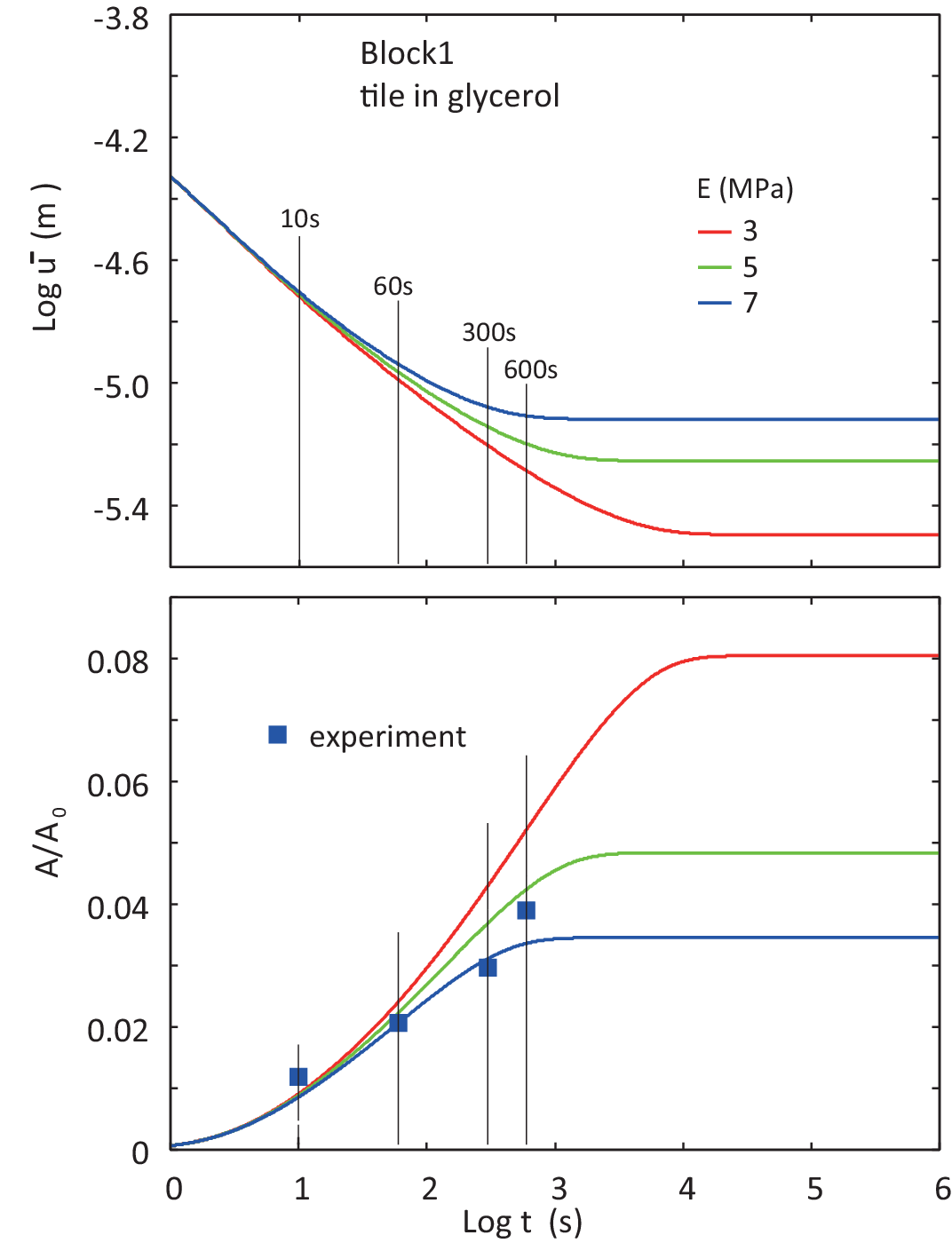}
\caption{\label{BLOCK.1logt.2Area.tile.eps}
The same as in Fig. \ref{BLOCK.1logt.2logu.RubberGlass.3.5.7MPa.eps}, but for
rubber squeezed against the tile surface. The squares in (b) show the relative contact areas
obtained from the experimental friction coefficients at the onset of sliding for different
waiting times in Fig. \ref{BLOCK.1distance.mu.tile.glycerol.Block1And4.eps}, multiplied by
the factor $p_0/\sigma_{\rm f}=0.06$. This corresponds to a frictional shear stress of
$\sigma_{\rm f}=p_0/0.06\approx2.7 \ {\rm MPa}$.
}
\end{figure}

The results in Figs. \ref{BLOCK.1x.2mu.tile.mud.1.and.4.eps},
\ref{BLOCK.1distance.mu.tile.glycerol.Block1And4.eps},
\ref{BLOCK.1slidingDistance.2mu.tile.grease.eps}, and
\ref{BLOCK.no.1.Block.no.4.smoothGLASSglycerol.eps} show that the friction force increases
monotonically with waiting time, i.e., with the duration of stationary contact before the onset
of sliding. We show below that, for the glycerol-lubricated tile and glass surfaces, the increase
in friction with waiting time is consistent with the theoretical predictions obtained using
(6), together with the relation between the mean surface separation $\bar u$ and the relative
contact area $A/A_0$ from Persson contact mechanics theory. In the low-pressure limit, the relation
between the asperity contact pressure and the mean surface separation is given by (8).
We assume that glycerol behaves as a Newtonian fluid with a room-temperature viscosity of
$\eta=1.4 \ {\rm Pa\,s}$.

Fig.~\ref{BLOCK.1logq.2logC.all.eps} shows the surface roughness power spectra as functions
of wavenumber on a log-log scale for the rubber surface (red) and the tile surface (blue).
The green curve is the sum of the power spectra of the two surfaces and therefore represents
the combined roughness spectrum of the rubber-tile interface. The thick curves are obtained
from measured surface topographies, whereas the thin curves represent linear extrapolations
outside the measured wavenumber range.

We assume that the friction force at the onset of sliding is proportional to the real contact area,
\[
F_{\rm f}=\sigma_{\rm f}A,
\]
where $\sigma_{\rm f}$ is the frictional shear stress acting within the real contact area $A$.
Since $F_{\rm N}=p_0A_0$, the friction coefficient is
\[
\mu={F_{\rm f}\over F_{\rm N}}
={\sigma_{\rm f}A\over p_0A_0}.
\]
Thus,
$${A\over A_0}=\mu{p_0\over\sigma_{\rm f}}.\eqno(11)$$

Fig.~\ref{BLOCK.1logt.2logu.RubberGlass.3.5.7MPa.eps} shows (a) the calculated mean surface
separation and (b) the relative contact area as functions of squeeze-out time for
Block1 squeezed against a flat glass surface in glycerol with viscosity
$\eta=1.4 \ {\rm Pa\,s}$. The red, green, and blue curves correspond to rubber Young's moduli
of $E=3 \ {\rm MPa}$, $5 \ {\rm MPa}$, and $7 \ {\rm MPa}$, respectively.
The squares in (b) show the relative contact areas obtained from (11), using the experimental
friction coefficients measured at the onset of sliding after the four different waiting times
shown in Fig. \ref{BLOCK.no.1.Block.no.4.smoothGLASSglycerol.eps}(a). In (11), we used
the nominal contact pressure $p_0=0.16 \ {\rm MPa}$ and a frictional shear stress of
$\sigma_{\rm f}=0.8 \ {\rm MPa}$.

Fig.~\ref{BLOCK.1logt.2Area.tile.eps} shows corresponding results for the tile surface.
In this case, a frictional shear stress of $\sigma_{\rm f}\approx2.7 \ {\rm MPa}$ was used.
The larger effective frictional shear stress for the tile surface may result from an additional
viscoelastic contribution to the friction force caused by deformation of the rubber by the
surface roughness of the tile.

There is reasonably good agreement between the theoretical predictions obtained using
$E=5 \ {\rm MPa}$ and the experimental data in
Figs. \ref{BLOCK.1logt.2logu.RubberGlass.3.5.7MPa.eps} and
\ref{BLOCK.1logt.2Area.tile.eps}. This agreement indicates that the theory provides an
accurate description of fluid squeeze-out during stationary contact.


\begin{figure}
\includegraphics[width=1.0\columnwidth]{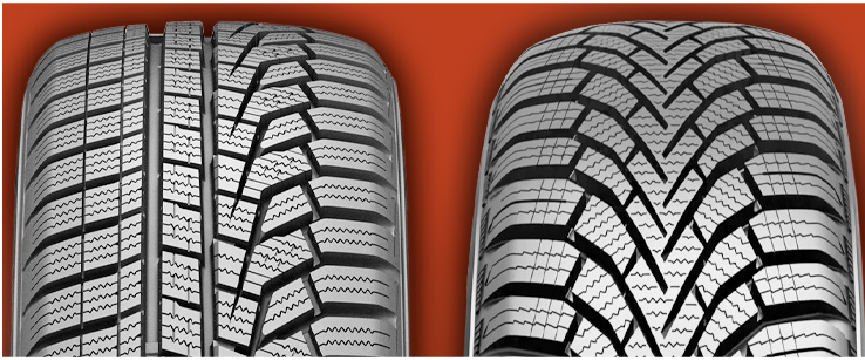}
\caption{\label{TreadPattern.ps}
Two tire-tread patterns. The large channels remove relatively thick fluid films, for which fluid
inertia can play an important role in the evacuation process. The smaller channels are important
for removing thin fluid films, for which viscous resistance controls the squeeze-out rate.
}
\end{figure}

\begin{figure}
\includegraphics[width=1.0\columnwidth]{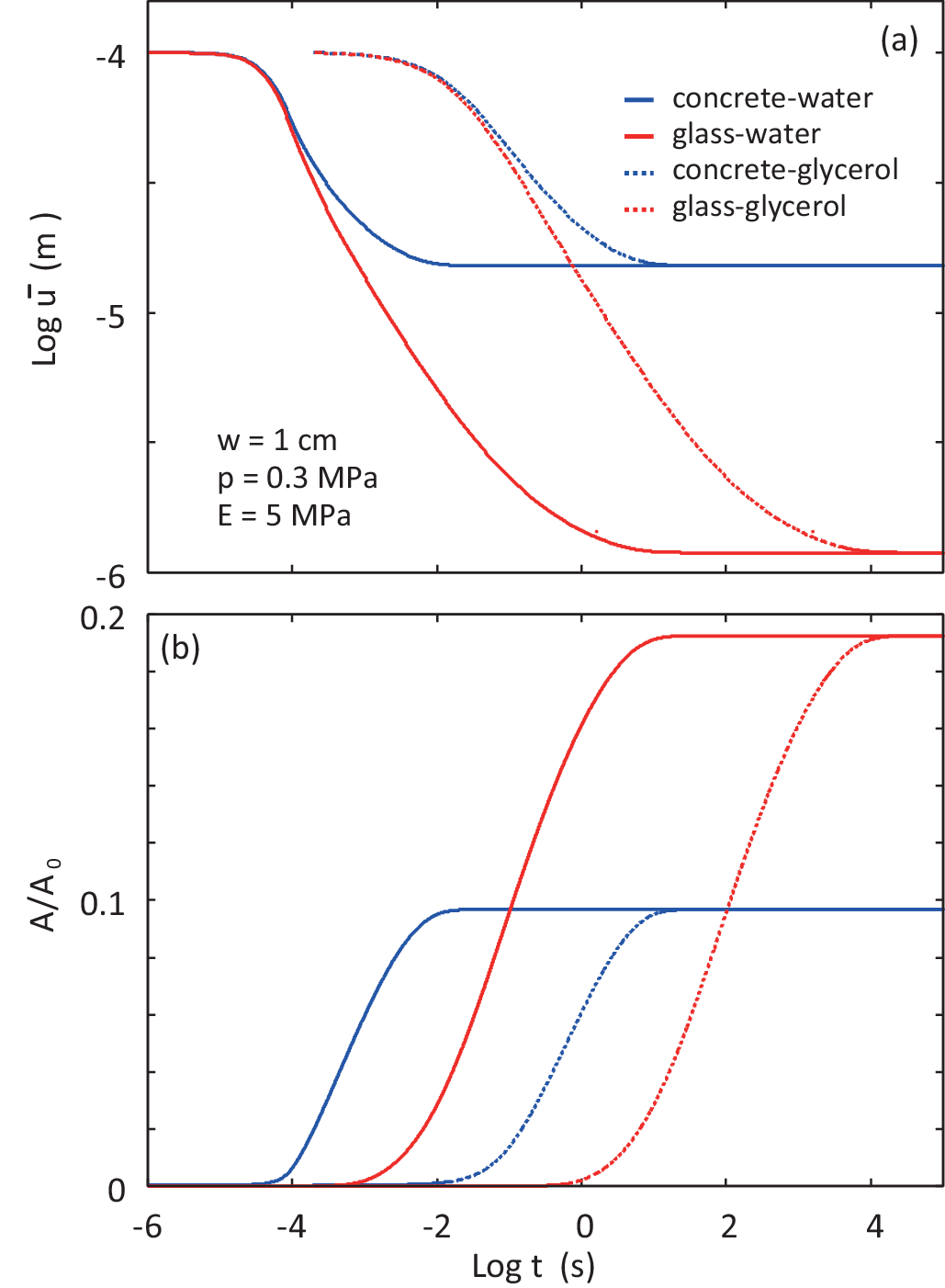}
\caption{\label{1Logt.2Area.and.Separation.Concrete.and.Rubber.eps}
The logarithm of the mean surface separation in (a) and the relative contact area in (b) for a
rubber block of width $w=1 \ {\rm cm}$ squeezed against a concrete surface (blue curves) or a
smooth glass surface (red curves). The solid curves correspond to water and the dashed curves
to glycerol, with assumed viscosities of $\eta=0.001 \ {\rm Pa\,s}$ and
$\eta=1 \ {\rm Pa\,s}$, respectively. The concrete surface has
$h_{\rm rms}=28 \ {\rm \mu m}$ and an rms slope of $0.74$, whereas the rubber surface has
$h_{\rm rms}=5.5 \ {\rm \mu m}$ and an rms slope of $0.37$. The glass surface is treated as
perfectly smooth.
}
\end{figure}

\vskip 0.1cm
\section{5 On the design for optimal fluid removal}

We first discuss the fluid squeeze-out mechanism illustrated in Fig. \ref{Scraping.eps}(a).
Passenger-car tires have networks of channels separating tread blocks of different shapes.
The exact shape of the tread blocks is generally less important than their characteristic size.
For rapid fluid removal, the distance from the center of a tread block to the nearest drainage channel
should be as small as possible, and this distance is controlled mainly by the tread-block dimensions.

The wide channels between the tread blocks of a tire, as illustrated in Fig. \ref{TreadPattern.ps},
facilitate the removal of relatively thick water films from the tire footprint. At high rolling or
sliding velocities, typically $v_0> 20 \ {\rm m/s}$ for passenger-car tires, resistance to water
removal is governed mainly by fluid inertia. The wide channels are therefore required to transport
large volumes of water. If most of the water cannot be removed during the available contact time,
inertial hydroplaning may occur, resulting in very low friction.

When the fluid film becomes sufficiently thin, viscous resistance rather than fluid inertia
controls the squeeze-out rate. If the contact time is too short, or the fluid viscosity is too
high, viscous hydroplaning may occur. To facilitate the removal of thin fluid films, tires often
contain dense arrays of narrow channels. Since only a small fluid volume must be removed when the
film is thin, even very narrow channels may provide sufficient drainage capacity. However, since
most road surfaces exhibit multiscale roughness, adding engineered channels at still finer length
scales will, in most cases, have only a minor additional effect on fluid removal.

Rubber sliding friction generally has two main contributions: a viscoelastic contribution caused by
the time-dependent deformation of the rubber by road asperities, and an adhesive contribution arising
from nanoscale regions of the rubber surface that undergo repeated cycles of attachment to the road,
stretching, and detachment \cite{Paris1}. If the fluid is not completely squeezed out and the surfaces
remain separated by a fluid film with a thickness of several nanometers or more, the viscoelastic
contribution may remain comparable to that on a dry surface, whereas the adhesive contribution is
strongly reduced or vanishes.

The relatively weak influence of the detailed tread-block shape on fluid removal is illustrated by
the large variety of tread patterns used for tires; see Fig. \ref{TreadPattern.ps}. If one particular
tread geometry were substantially more effective than all others, one would expect that geometry to
dominate tire design, which is not observed.

Most vehicles are equipped with anti-lock braking systems, and during braking the tires roll with a
finite but generally small slip. If the tire-road footprint has a length $d=10 \ {\rm cm}$ in the
rolling direction and the vehicle velocity is $v_{\rm c}=30 \ {\rm m/s}$, corresponding to
$108 \ {\rm km/h}$, then the rolling velocity is approximately $v_{\rm R}\approx v_{\rm c}$.
A tread block therefore remains within the footprint for a time
$\Delta t\approx d/v_{\rm R}\approx3\times10^{-3} \ {\rm s}$. A shorter time scale of interest is
approximately $3\times10^{-4} \ {\rm s}$, corresponding to the time required for the tread block
to travel $1 \ {\rm cm}$ within the footprint.

Fig.~\ref{1Logt.2Area.and.Separation.Concrete.and.Rubber.eps} illustrates fluid squeeze-out for
four cases: water, shown by the solid curves, and glycerol, shown by the dashed curves, on a concrete
surface and on a smooth glass surface. In the calculations, we used the measured surface roughness
power spectrum of a concrete surface with rms roughness
$h_{\rm rms}=28 \ {\rm \mu m}$ and rms slope $0.74$. The glass surface was treated as perfectly smooth,
so that only the roughness of the rubber surface was included. The rubber surface had
$h_{\rm rms}=5.5 \ {\rm \mu m}$ and an rms slope of $0.37$.

The rubber block was assumed to have a width $w=1 \ {\rm cm}$ in the $x$ direction and to be infinitely
long in the $y$ direction. We used a rubber Young's modulus of $E=5 \ {\rm MPa}$ and a nominal contact
pressure of $0.3 \ {\rm MPa}$, representative of tire applications.

Fig.~\ref{1Logt.2Area.and.Separation.Concrete.and.Rubber.eps}(a) shows the mean surface separation,
and Fig. \ref{1Logt.2Area.and.Separation.Concrete.and.Rubber.eps}(b) shows the relative contact area,
both as functions of the logarithm of time. These results can be used to estimate how the relative
contact area evolves as a tread block moves from the leading edge to the trailing edge of the footprint.

\begin{figure}
\includegraphics[width=1.0\columnwidth]{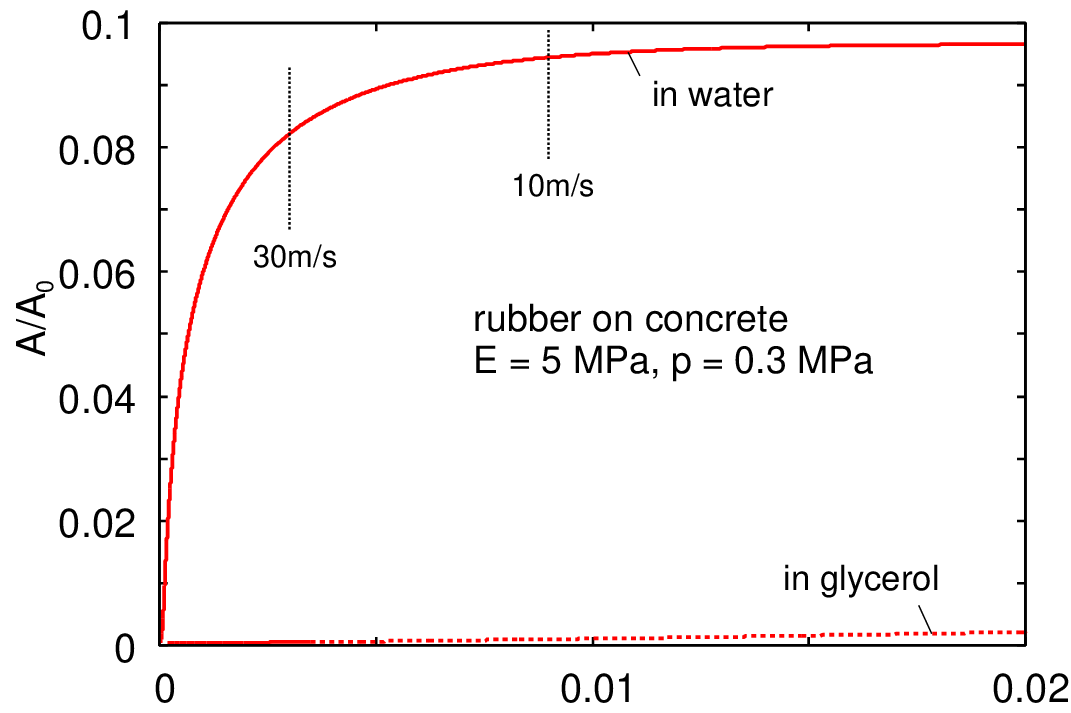}
\caption{\label{1time.2Area.concrete.water.glycerol.tire.eps}
The same results as in Fig. \ref{1Logt.2Area.and.Separation.Concrete.and.Rubber.eps} for a concrete
surface lubricated with water or glycerol. The vertical dotted lines indicate the residence times of
a tread block in a $10 \ {\rm cm}$ long footprint at rolling velocities of
$10 \ {\rm m/s}$ and $30 \ {\rm m/s}$.
}
\end{figure}

Fig.~\ref{1time.2Area.concrete.water.glycerol.tire.eps} shows the same results for the concrete
surface lubricated with water or glycerol. The two vertical dotted lines indicate the residence time
of a tread block in a footprint of length $10 \ {\rm cm}$ at rolling velocities of
$10 \ {\rm m/s}$ and $30 \ {\rm m/s}$. For the water-lubricated surface at
$v_{\rm R}=30 \ {\rm m/s}$, the real contact area reaches approximately $85\%$ of its final
complete-squeeze-out value by the time the tread block leaves the footprint. More importantly, it
already reaches approximately $75\%$ of its final value when the tread block arrives at the center
of the footprint.

\begin{figure}
\includegraphics[width=1.0\columnwidth]{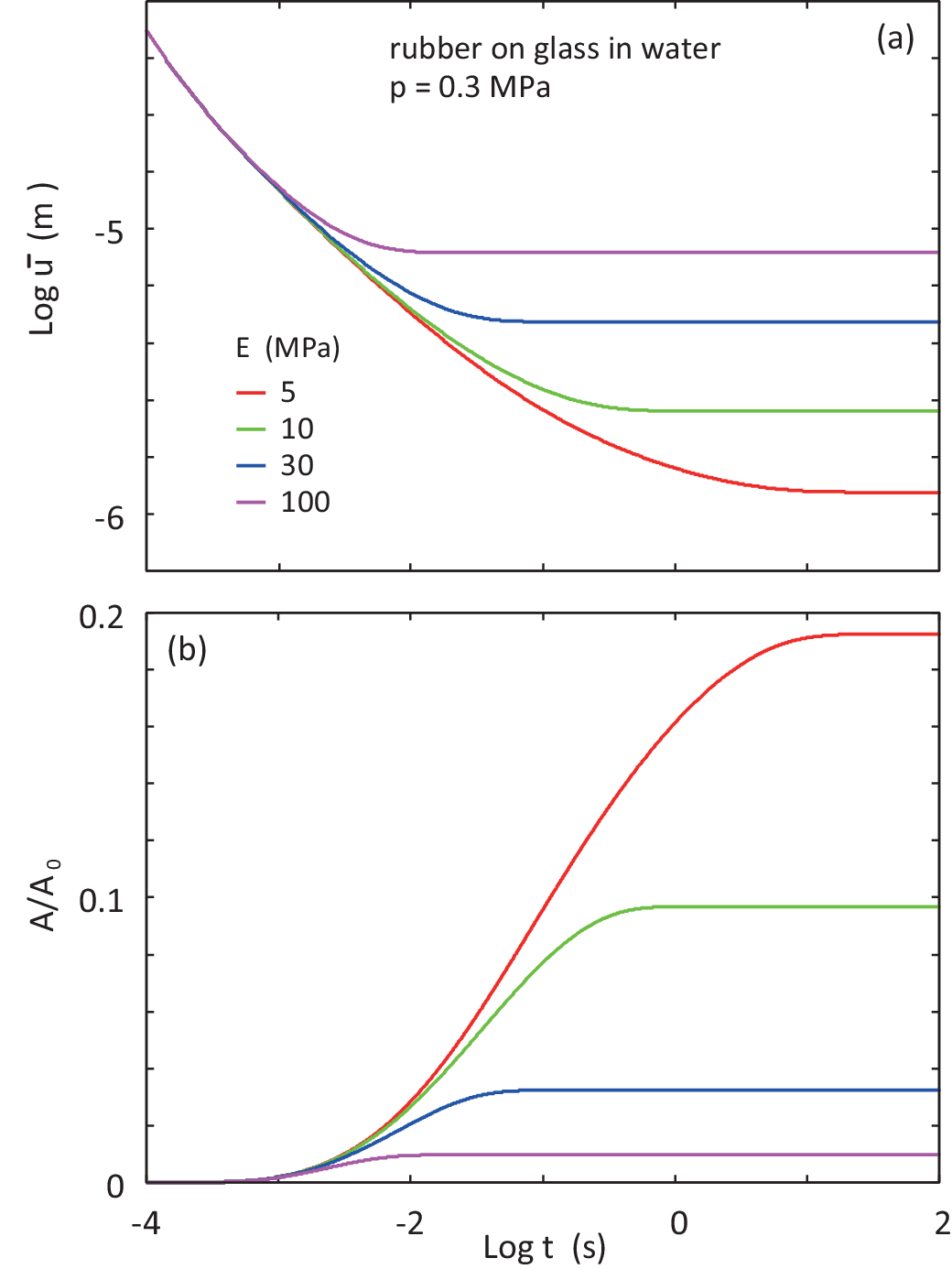}
\caption{\label{1logt.2Logu.2Area.Shoe.Rubber.p5.10.30.100GPa.eps}
The logarithm of the mean surface separation in (a) and the relative contact area in (b) for a
rubber block of width $w=1 \ {\rm cm}$ squeezed against a smooth glass surface in water with
viscosity $\eta=0.001 \ {\rm Pa\,s}$, for different values of the rubber Young's modulus.
The rubber surface has rms roughness $h_{\rm rms}=5.5 \ {\rm \mu m}$ and rms slope $0.37$.
The glass surface is treated as perfectly smooth.
}
\end{figure}

We next investigate how fluid squeeze-out depends on the elastic modulus $E$ of the rubber.
Before asperity contact occurs, changing $E$ has little influence on the squeeze-out process.
At long times, however, increasing $E$ results in a larger mean surface separation and a smaller
relative contact area. This is illustrated in
Fig. \ref{1logt.2Logu.2Area.Shoe.Rubber.p5.10.30.100GPa.eps} for a rubber block squeezed against
a glass surface in water, using $E=5$, $10$, $30$, and $100 \ {\rm MPa}$.

The low-frequency rubbery modulus can be increased by increasing either the filler concentration
or the cross-link density. In addition, the effective modulus depends on deformation frequency
$\omega$ and temperature. For a rolling tire, a characteristic deformation frequency is
$\omega\sim v_{\rm R}/d$, where $d$ is of the order of the tire-road footprint length, or somewhat
smaller. For footwear, the characteristic frequency is approximately $\omega\sim1/t^*$, where
$t^*$ is the duration of contact between the foot and the ground.

For tires, this estimate typically gives $\omega\approx100$-$1000 \ {\rm s}^{-1}$, whereas for
shoes during walking it gives $\omega\approx1 \ {\rm s}^{-1}$. Since $E(\omega)$ is complex at
nonzero frequency, the effective modulus used in estimating squeeze-out may be approximated by
$E_{\rm eff}=|E(\omega)|$. Since the strain in the asperity contact regions may be large, an
effective large-strain modulus should be used, typically corresponding to strains of the order of
$\epsilon\approx0.5$.

\begin{figure}
\includegraphics[width=0.9\columnwidth]{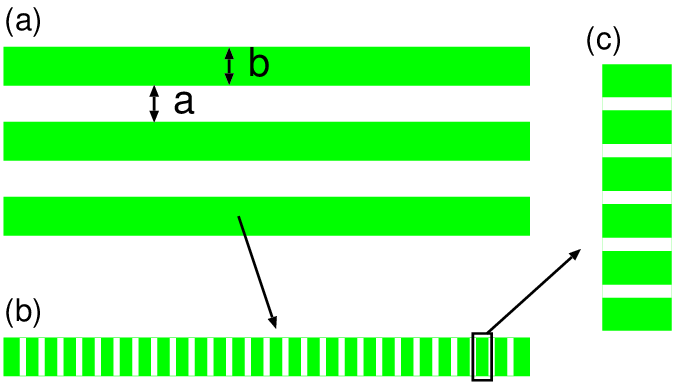}
\caption{\label{Linear.eps}
(a) Rectangular rubber blocks occupying a fraction $b/a$ of the nominal surface area.
The fluid squeeze-out rate increases when narrower channels are added, as in (b), and increases
further when still narrower channels are introduced, as in (c).
}
\end{figure}

In some applications, fluid must be squeezed out from between rubber tread blocks and very smooth
substrates. If both the rubber and the substrate are very smooth, squeeze-out can be
strongly accelerated by using a hierarchical network of channels distributed over three or more
length scales, as illustrated in Fig. \ref{Linear.eps}.

Fig.~\ref{Linear.eps}(a) shows rectangular tread blocks that are narrow in the fluid-flow direction
and wide in the transverse direction. In Fig. \ref{Linear.eps}(b), additional narrower channels are
introduced into the tread blocks, accelerating fluid removal in the regime where viscous resistance
dominates. More generally, a hierarchical construction in which large tread blocks are separated by
wide channels, subdivided by a network of finer channels, and then further subdivided by still finer
channels can strongly accelerate squeeze-out between very smooth surfaces in the viscous-flow regime.

A related principle is used by tree frogs to adhere to water-flooded surfaces. Tree frogs adhere to dry
surfaces at least partly through capillary adhesion involving secreted fluid. On flooded surfaces,
adhesion may instead involve suction, but this first requires removal of most of the fluid from between
the adhesive pads and the substrate.

\begin{figure}
\includegraphics[width=1.0\columnwidth]{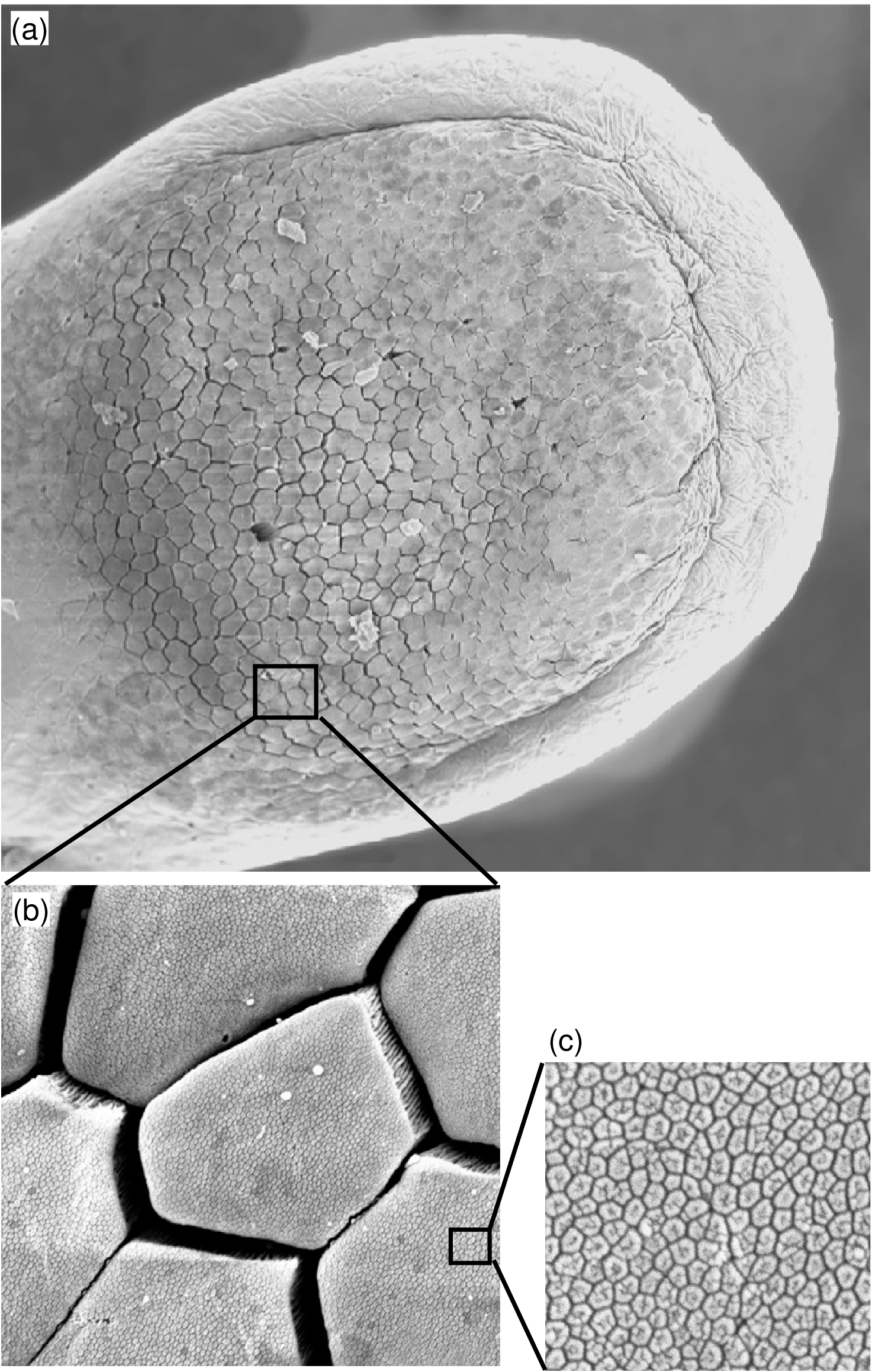}
\caption{\label{TreeFogPadPic.eps}
The adhesive pads on the toes of a tree frog have diameters of approximately $1 \ {\rm mm}$ and
consist mainly of hexagonal cells with diameters of approximately $10 \ {\rm \mu m}$, separated by
channels approximately $1 \ {\rm \mu m}$ wide. The surfaces of the hexagonal cells are covered with
nanoscale peg-like protrusions, referred to as nanopillars. The nanopillars are prismatic structures
separated by a nanoscale channel network analogous to the microscale channel network between the
hexagonal cells \cite{Federle}.
}
\end{figure}

Fig.~\ref{TreeFogPadPic.eps} shows the adhesive pad of a tree frog. The pad diameter is approximately
$1 \ {\rm mm}$ and the surface consists mainly of hexagonal cells with diameters of approximately
$10 \ {\rm \mu m}$, separated by channels approximately $1 \ {\rm \mu m}$ wide. The surfaces of the
hexagonal cells are covered with nanoscale peg-like protrusions, commonly referred to as nanopillars.
These nanopillars are prismatic structures separated by a nanoscale channel network analogous to the
microscale channel network between the hexagonal cells \cite{Federle,frodo1,frodo2}.

If the adhesive pad on a tree-frog toe is regarded as analogous to a tire tread block, fluid squeeze-out
occurs over three characteristic length scales. At large film thicknesses, fluid is removed at the
scale of the entire pad. At intermediate film thicknesses, drainage occurs through the larger channels
between the hexagonal cells. Finally, for nanometer-scale films, fluid is removed through the nanoscale
channels between the nanopillars.

We have argued that, in the viscous-flow regime, the squeeze-out rate increases as the tread-block
size decreases. One may therefore ask why tire tread blocks are not made extremely small, for example
of micrometer dimensions as in tree-frog adhesive pads. The principal limitations are wear resistance
and tread-block stability.

Since rubber wear gradually removes material, the channels separating the tread blocks must be
sufficiently deep that they do not disappear after a short period of tire use. However, if the
cross-sectional dimensions of a tread block are much smaller than the depth of the surrounding
channels, the block behaves as a compliant fiber and may bend strongly under normal and tangential
loading.

\begin{figure}
\includegraphics[width=1.0\columnwidth]{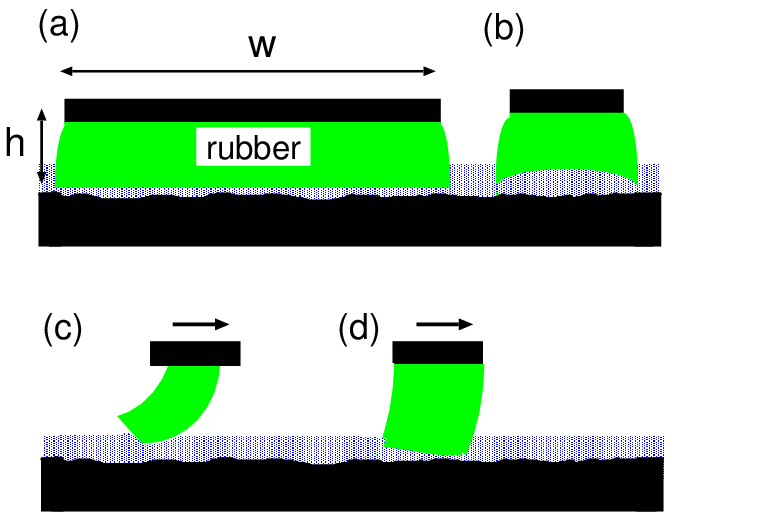}
\caption{\label{BENDING.eps}
A rubber block (width $w$ and height $h_0$) bonded to a rigid plate at its upper surface and squeezed against a rigid substrate
in a fluid. If $w/h_0\gg1$, the lower surface remains nearly flat and the dynamic scraping process
in Fig. \ref{Scraping.eps}(b) is ineffective. When $w/h_0\approx1$, this mechanism can become
effective. If $w/h_0<1$, an isolated rubber block may bend excessively, promoting hydrodynamic
lubrication and reducing its scraping efficiency. The optimal degree of bending is not known to us, but
an appropriate deformation should generate negative fluid pressure behind the leading edge,
thereby enhancing scraping.
}
\end{figure}

\begin{figure}
\includegraphics[width=0.47\textwidth,angle=0.0]{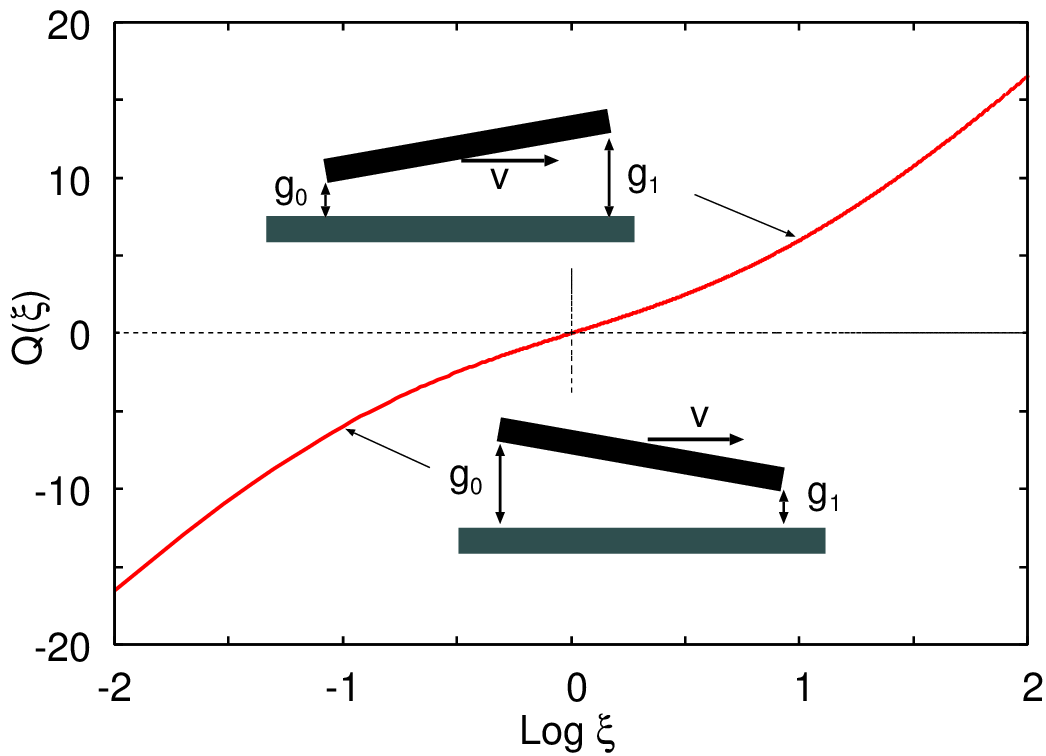}
\caption{\label{Q1xi.2pressure.tilted.junction.eps}
The function $Q(\xi)$ as a function of the logarithm of the gap ratio $\xi=g_1/g_0$.
The function is antisymmetric under inversion, $Q(\xi)=-Q(1/\xi)$.
}
\end{figure}

We now consider the two scraping processes illustrated in Fig. \ref{Scraping.eps}(b) and (c).
Consider first a rectangular block (length $L$ and width $w$) sliding at velocity $v_0$ on a flat substrate in a fluid.
Assume that the lower surface is tilted, as indicated in the inset of
Fig. \ref{Q1xi.2pressure.tilted.junction.eps}, with gaps $g_1$ and $g_0$ at the entrance and exit,
respectively. If the block is much longer in the direction perpendicular to sliding than in the
sliding direction, $w \ll L$, the normal force per unit length exerted by the fluid on the block is
$$
{F_{\rm N} \over L} ={\eta w^2v_0\over(g_0+g_1)^2}Q(\xi),
$$
where $w$ is the width in the sliding direction and
$$
Q(\xi)=6\left({\xi+1\over\xi-1}\right)^2
\left[\ln\xi-2{\xi-1\over\xi+1}\right],
\qquad
\xi={g_1\over g_0}.
\eqno(12)
$$

Fig.~\ref{Q1xi.2pressure.tilted.junction.eps} shows $Q(\xi)$ as a function of
$\log\xi$. It follows directly from (12) that
$Q(1/\xi)=-Q(\xi)$, as is also evident from the figure. Hence, if the lower surface is tilted such
that the gap at the entrance is smaller than that at the exit, the hydrodynamic pressure is negative,
so that $F_{\rm N}<0$ and the two surfaces are pulled together during sliding.

Equation (12) predicts that the magnitude of $F_{\rm N}$ increases without bound, approximately as
$\ln\xi$, as $\xi$ becomes either very large or very small. However, the lubrication approximation
used to derive (12) assumes that both $g_0$ and $g_1$ are much smaller than $w$. This approximation
breaks down for large surface inclination. In the limiting case where
$|g_1-g_0|\sim w$, the surface approaches a vertical orientation, and the hydrodynamic force should
act predominantly parallel to the substrate, implying that the normal force tends toward zero.
Consequently, $|F_{\rm N}|$ must attain a maximum at some intermediate tilt angle between
$0$ and $90^\circ$.

Tilted surface regions occur in both scraping processes illustrated in
Fig. \ref{Scraping.eps}(b) and (c). In the configuration shown in
Fig. \ref{Scraping.eps}(b), negative pressure near the leading edge tends to close the contact,
whereas positive pressure near the trailing edge tends to open it. This pressure distribution
promotes removal of fluid from the interface.

For the configuration shown in Fig. \ref{Scraping.eps}(c), negative pressure develops behind the
leading edge. This pressure adds to the externally applied normal load, increases the local contact
pressure between the rubber and the substrate near the leading edge, and thereby enhances
scraping of the fluid.

Assume that the rubber block is bonded to a rigid plate at its upper surface and that friction
between the lower surface of the block and the substrate can initially be neglected, as may be
expected when the solids are separated by a fluid film. For the dynamic scraping process shown in
Fig. \ref{Scraping.eps}(b) to be important, the following conditions should be satisfied.

(a) The width of the block in the sliding direction should not be much larger than its height $h_0$
or else it will be effectively very stiff. This is best illustrated with a
circular cylinder or disc with the radius $R$.
If the disc is bound to a flat rigid surface on one side, the rubber becomes effectively very stiff
with respect to compressive deformations varying over a lateral length scale of the order of $R$.
This problem has been studied in detail by Gent \cite{Gent1,Gent2,Gent3,Gent4}, and a simpler derivation 
was presented in Ref.~\cite{PerssonG}. During compression of a bonded rubber
disc under a pressure $p$, the strain $\epsilon=(h_0-h)/h_0$ satisfies
$p=E_{\rm eff}\epsilon$ where
$$
E_{\rm eff}\approx
\left[
1+{1\over8}\left({R\over h_0}\right)^2
\right]E.
$$

If $R>3h_0$, the effective modulus may be much larger than the Young's modulus $E$ measured for a
rubber strip in compression. In this case, when the rubber block is squeezed against a substrate
in a fluid, the lower surface undergoes little bending
[see Fig. \ref{BENDING.eps}(a)], and the dynamic scraping mechanism shown in
Fig. \ref{Scraping.eps}(b) becomes ineffective.

This increase in effective stiffness applies only to deformations with lateral wavelengths of the
order of the block thickness or larger. For deformation of the rubber by surface roughness with
wavelengths much smaller than $h_0$, the ordinary bulk modulus $E$ remains the appropriate modulus.

(b) The maximum transient bending of the lower surface increases with increasing applied squeezing
pressure.

(c) Sliding should begin relatively soon after application of the normal force, when transient
bending of the lower surface is near its maximum. At long times, as more fluid is removed through
the squeeze-out process in Fig. \ref{Scraping.eps}(a), the bending deformation of the lower surface decreases.

(d) The mechanism is most important for highly viscous fluids, mud, or grease. In these cases,
the stationary squeeze-out through process (a) in Fig. \ref{Scraping.eps} is slow, and the system remains
for a long time in a state where the lower surface is bent, as illustrated in
Fig. \ref{BENDING.eps}(b). Dynamic scraping is particularly important for such systems because
stationary squeeze-out alone may require a very long time to remove most of the fluid.

(e) Softer rubber generally produces greater bending and therefore stronger dynamic scraping.
This behavior was observed in Ref. \cite{JCPsqueeze}, where rubber blocks sliding on surfaces
covered with grease or glycerol reached their final friction state after a shorter sliding distance
when a softer rubber compound was used.

We next consider the scraping process shown in Fig. \ref{Scraping.eps}(c). This mechanism also places
constraints on the geometry of the rubber block. If the height-to-width ratio $h_0/w$ is too large,
or if the rubber is too soft, the block may bend excessively, as illustrated in
Fig. \ref{BENDING.eps}. This assumes that the channels between adjacent blocks are sufficiently wide
to permit the deformation. Excessive bending may promote hydrodynamic lift rather than effective
scraping.

Conversely, if $w\gg h_0$, the block bends very little and deforms mainly in shear, resulting in weak
scraping. The optimal value of $w/h_0$ is not known to us 
and will depend on the elastic properties of the
rubber and on the loading and lubrication conditions.

\begin{figure}
\includegraphics[width=0.47\textwidth,angle=0.0]{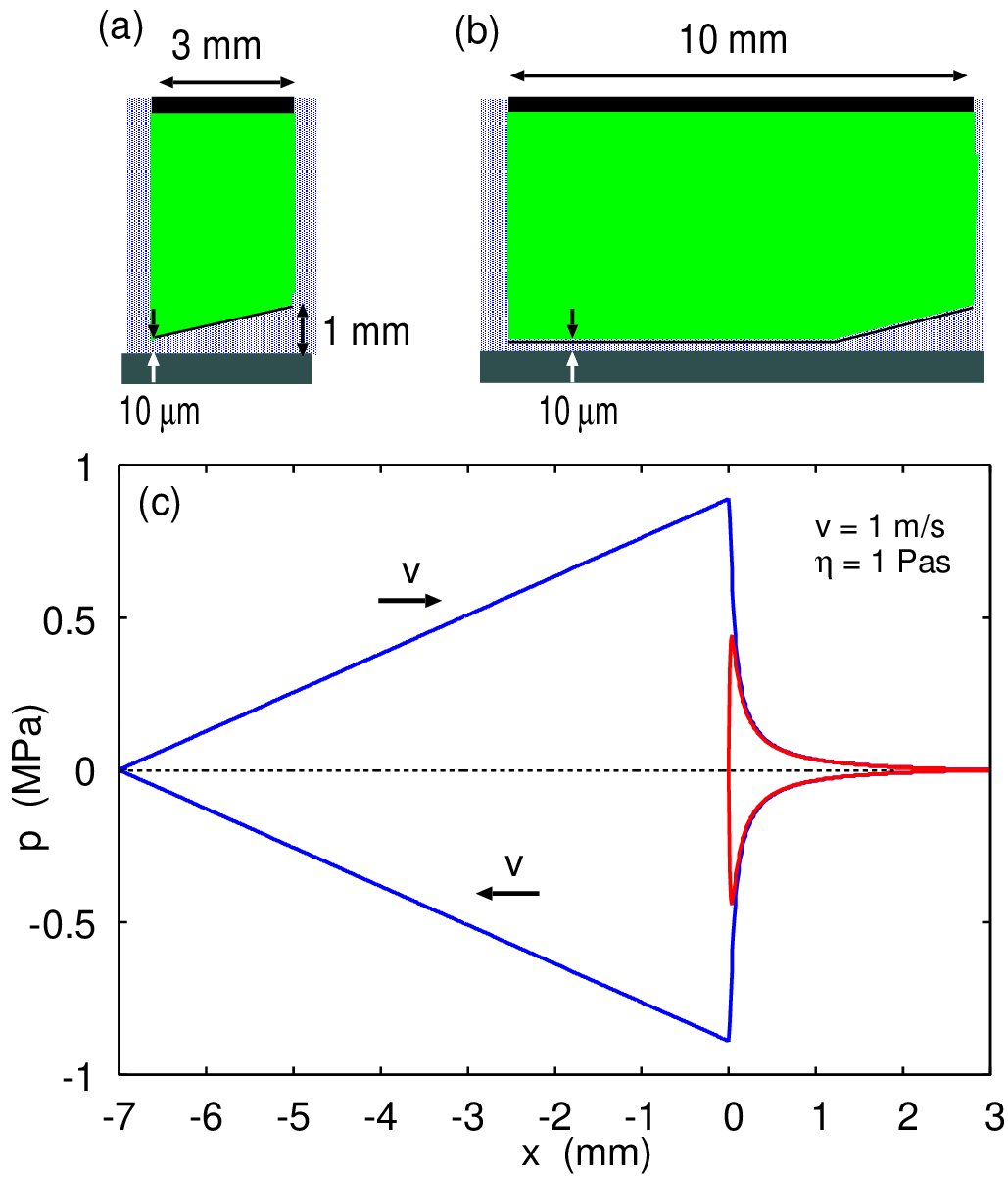}
\caption{\label{TILTED.japan.1x.2pressure.eps}
The fluid pressure beneath a block with a tilted lower surface. The pressure is positive when the
block slides in the direction for which fluid enters from the side with the larger gap and negative
when it slides in the opposite direction, as shown by the $p>0$ and $p<0$ curves in (c).
The red curves in (c) correspond to the block shown in (a). Extending the block with a flat region
on the side having the smaller gap, as shown in (b), increases the magnitude of the pressure, as
indicated by the blue curves in (c). The calculations use the block dimensions shown in the figure,
sliding velocity $v=1 \ {\rm m/s}$, and viscosity $\eta=1 \ {\rm Pa\,s}$. Results for other values
of $v$ and $\eta$ follow from the scaling relation $p\sim\eta v$.
}
\end{figure}

If the rubber block is much wider than it is high, spontaneous bending of the block may be negligible.
Nevertheless, the block can be designed with an intentionally tilted lower surface, as illustrated
in Fig. \ref{TILTED.japan.1x.2pressure.eps}(a) and (b). For such a surface, negative fluid pressure
develops when the block moves in the direction for which fluid is entrained into the narrowing gap.

Fig.~\ref{TILTED.japan.1x.2pressure.eps}(c) shows the calculated fluid-pressure distributions.
The pressure is positive when fluid enters from the side with the larger gap and negative when the
sliding direction is reversed. The red curves correspond to the block geometry shown in
Fig. \ref{TILTED.japan.1x.2pressure.eps}(a). Adding a flat section on the side with the smaller gap,
as shown in Fig. \ref{TILTED.japan.1x.2pressure.eps}(b), increases the magnitude of the generated
pressure, as illustrated by the blue curves.

The calculations were performed for the dimensions indicated in the figure, using
$v=1 \ {\rm m/s}$ and $\eta=1 \ {\rm Pa\,s}$. The pressure at other velocities and viscosities can
be obtained from the scaling relation $p\sim\eta v$.

If the fluid is saturated with atmospheric gases, meaning that it is in equilibrium with the
atmosphere at ambient pressure, dissolved gas may come out of solution when the local absolute
pressure is reduced sufficiently below the ambient pressure. Even in the absence of dissolved
gases, vapor cavitation can occur when the absolute pressure falls below the fluid vapor pressure,
which is approximately $3 \ {\rm kPa}$ for water at room temperature. At an ambient pressure of
approximately $100 \ {\rm kPa}$, this corresponds to a gauge pressure of approximately
$-97 \ {\rm kPa}$, or nearly one atmosphere below ambient pressure. However, cavitation requires
nucleation, and moderate negative pressures may persist if suitable nucleation sites are
absent. Negative pressures may also persist when the fluid is undersaturated with dissolved
atmospheric gases.

Such negative fluid pressures were observed by Ishizako et al.\ for silicone-rubber sliders moving
on smooth glass surfaces lubricated with glycerol \cite{Japan3,Japan4}. In some of these systems,
pressures as low as approximately $-50 \ {\rm kPa}$ relative to atmospheric pressure were
reported.

\section{6 Summary and conclusions}

We have studied the sliding friction of rectangular rubber blocks against tile and glass surfaces
under both dry and lubricated conditions. Particular attention was given to the dependence of the friction
force on the waiting time before the onset of sliding and to the relative importance of stationary
squeeze-out and sliding-induced fluid removal.

For rectangular rubber blocks sliding on tile surfaces covered with mud or grease, the steady-sliding
state is reached after a sliding distance of the order of a few times the block dimension in the
sliding direction, essentially {\it independent} of the preceding waiting time. This demonstrates
that fluid removal is governed mainly by the scraping processes illustrated in
Fig.~\ref{Scraping.eps}(b) and (c). For contacts lubricated with glycerol, both stationary squeeze-out and scraping,
corresponding to process (a)-(c) in Fig.~\ref{Scraping.eps}, contribute
significantly to fluid removal. For water, the low viscosity generally allows squeeze-out to be
nearly complete before the onset of sliding.

For low-viscosity fluids such as water, stationary squeeze-out through mechanism (a) in
Fig.~\ref{Scraping.eps} is likely to dominate in many cases. For highly viscous substances such as
mud and grease, however, the present results and those reported in Ref.~\cite{JCPsqueeze} indicate
that some finite slip is often required for efficient fluid removal. Under these conditions, the scraping
processes shown in Fig.~\ref{Scraping.eps}(b) and (c) become particularly important.

These results suggest that rapid fluid removal requires different design strategies at different
film thicknesses. Wide channels are effective for evacuating thick fluid films, whereas smaller
and more densely distributed channels reduce the drainage distance in the viscous-flow regime.
However, for road surfaces with multiscale roughness, adding engineered channels at still finer
length scales is expected to provide only limited additional benefit. Dynamic scraping can be
enhanced by selecting block dimensions and rubber stiffness that permit sufficient transient
bending without compromising tread-block stability. Intentionally tilted lower surfaces may
provide an additional means of generating negative fluid pressure near the leading edge and
promoting contact closure.

\vskip 0.3cm
\section{Appendix: Rubber friction on biofilm-contaminated glass}

We have
 performed rubber-friction experiments on two glass surfaces contaminated with biofilms.
The glass surfaces were cleaned with isopropanol, and one of them was subsequently sandblasted.
Both surfaces were immersed in a lake for $10$ days, during which biofilms formed on the surfaces.
The friction experiments were performed on the wet, biofilm-contaminated surfaces immediately after
they were removed from the lake.

\begin{figure}
\includegraphics[width=1.0\columnwidth]{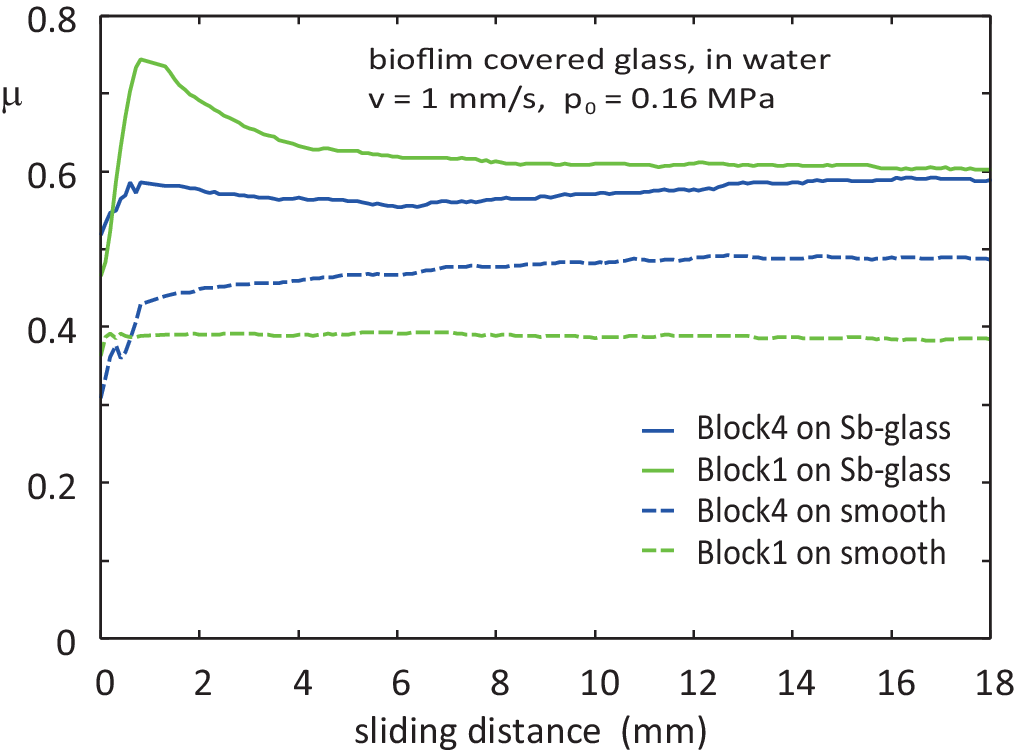}
\caption{\label{BLOCK.distance.2mu.contaminated.glass.eps}
The friction coefficient as a function of sliding distance for Block1 (green curves)
and Block4 (blue curves) on the smooth biofilm-contaminated glass surface (dashed curves)
and the sandblasted biofilm-contaminated glass surface (solid curves). The sliding speed is
$v=1 \ {\rm mm/s}$, and the nominal pressure is $p_0=0.16 \ {\rm MPa}$.
}
\end{figure}

Fig.~\ref{BLOCK.distance.2mu.contaminated.glass.eps} shows the friction coefficient as a function
of sliding distance for Block1, represented by the green curves, and Block4,
represented by the blue curves. The dashed curves correspond to the smooth contaminated glass
surface, whereas the solid curves correspond to the sandblasted contaminated surface. As expected,
the sandblasted glass surface exhibits higher friction. In all cases, the friction increases during
approximately the first $1 \ {\rm mm}$ of sliding.

\begin{figure}
\includegraphics[width=1.0\columnwidth]{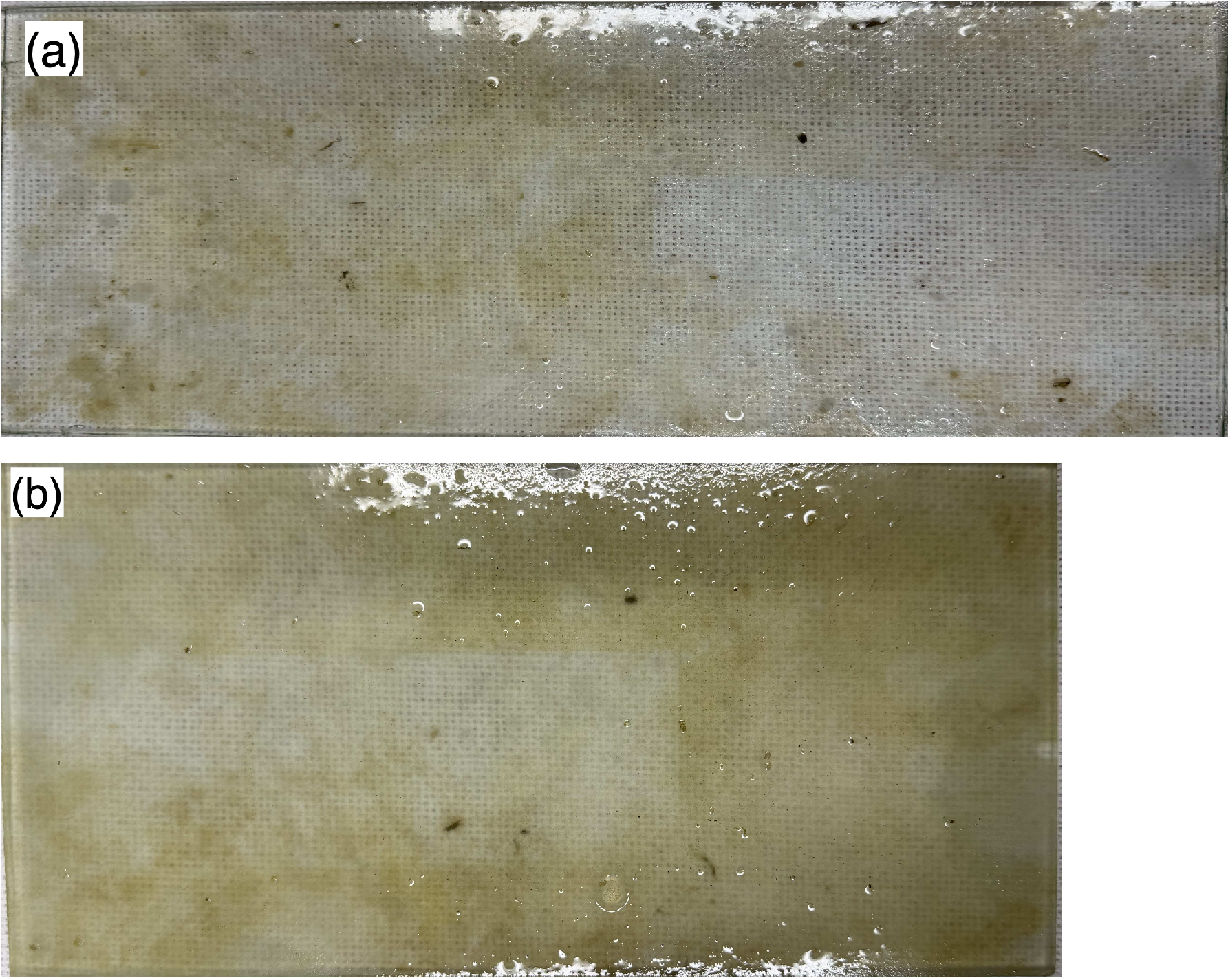}
\caption{\label{GlassPondPicture.eps}
Optical images of the (a) smooth and (b) sandblasted glass surfaces after immersion in lake water.
The sliding tracks are visible as regions from which the biofilm contamination was partially removed.
}
\end{figure}

Fig.~\ref{GlassPondPicture.eps} shows optical images of the glass plates after the sliding-friction
experiments. Some removal of the biofilm during sliding can be observed, although it is not very
obvious in the images because both sides of the plates are covered with biofilm.

\end{document}